\documentclass[twocolumn,showpacs,superscriptaddress,preprintnumbers,prd]{revtex4}
\usepackage{graphicx,amsmath,dcolumn,bm}
\begin{document}
\preprint{KEK-TH-1108}
\title{\Large \bf Determination of fragmentation functions
                  and their uncertainties}
\author{M. Hirai}
\email[E-mail: ]{mhirai@th.phys.titech.ac.jp}
\affiliation{Department of Physics,
             Tokyo Institute of Technology \\
             Ookayama, Meguro-ku, Tokyo, 152-8550, Japan}
\author{S. Kumano}
\email[E-mail: ]{shunzo.kumano@kek.jp}
\affiliation{Institute of Particle and Nuclear Studies \\
          High Energy Accelerator Research Organization (KEK) \\
          1-1, Ooho, Tsukuba, Ibaraki, 305-0801, Japan}
\affiliation{Department of Particle and Nuclear Studies \\
             Graduate University for Advanced Studies \\
           1-1, Ooho, Tsukuba, Ibaraki, 305-0801, Japan}     
\author{T.-H. Nagai}
\email[E-mail: ]{tnagai@post.kek.jp}
\affiliation{Department of Particle and Nuclear Studies \\
             Graduate University for Advanced Studies \\
           1-1, Ooho, Tsukuba, Ibaraki, 305-0801, Japan}      
\author{K. Sudoh}
\email[E-mail: ]{kazutaka.sudoh@kek.jp}
\affiliation{Institute of Particle and Nuclear Studies \\
          High Energy Accelerator Research Organization (KEK) \\
          1-1, Ooho, Tsukuba, Ibaraki, 305-0801, Japan}
\homepage[URL: ]{http://research.kek.jp/people/kumanos/ffs.html}
\date{\today}
\begin{abstract}
Fragmentation functions and their uncertainties are determined for pion, kaon,
and proton by a global $\chi^2$ analysis of charged-hadron production
data in electron-positron annihilation and by the Hessian method
for error estimation. It is especially important that the uncertainties of
the fragmentation functions are estimated in this analysis. The results
indicate that the fragmentation functions, especially gluon and
light-quark fragmentation functions, have large uncertainties
at small $Q^2$. There are large differences between widely-used functions
by KKP (Kniehl, Kramer, and P\"otter) and Kretzer; however, they
are compatible with each other and also with our functions 
if the uncertainties are taken into account. We find that
determination of the fragmentation functions is improved in
next-to-leading-order (NLO) analyses for the pion and kaon in comparison
with leading-order ones. Such a NLO improvement is not obvious
in the proton. Since the uncertainties are large at small $Q^2$,
the uncertainty estimation is very important for analyzing hadron-production
data at small $Q^2$ or $p_T$ ($Q^2, \, p_T^2 << M_Z^2$) in lepton scattering
and hadron-hadron collisions. A code is available for general users
for calculating obtained fragmentation functions.
\end{abstract}

\maketitle

\section{Introduction}

In finding any exotic physics signatures beyond the standard model
and any new hadronic systems in high-energy hadron reactions, 
for example at RHIC (Relativistic Heavy Ion Collider) and 
LHC (Large Hadron Collider), it is necessary to have accurate
QCD (Quantum Chromodynamics) predictions for cross sections. 
The perturbative QCD is now understood in the next-to-leading-order (NLO)
level for many reactions and in the next-to-next-to-leading order (NNLO)
for some processes. Therefore, the crucial part is to understand
nonperturbative aspects, namely parton distribution functions (PDFs)
and fragmentation functions. 

Fragmentation functions are used in high-energy reaction processes
with hadron production. They include hadron-production processes
in electron-positron annihilation, electron, muon, and neutrino scattering
from proton or nucleus, proton-proton collisions, and heavy-ion
collisions \cite{sidis,rhic-pi}. 
Such reactions are becoming increasingly important in hadron
physics for investigating the origin of the proton spin and quark-hadron
matters in heavy-ion reactions.

The fragmentation functions are related to a nonperturbative aspect
of QCD, so that they cannot be precisely calculated by theoretical
methods at this stage.
The situation is similar to the determination of the PDFs, where
high-energy experimental data are used for their determination
instead of theoretical calculations.
There are already several studies on such analyses \cite{ff-web}.
Widely used parametrizations were obtained by Kniehl, Kramer, and
P\"otter (KKP) \cite{kkp} and also by Kretzer \cite{kretzer}.
An updated version was reported recently by Albino, Kniehl, and
Kramer (AKK) \cite{kkp}. The fragmentation functions were
determined by analyzing the hadron-production data in the $e^+e^-$
annihilation. The analysis was done in leading order (LO) and
NLO of running coupling constant $\alpha_s$.

Despite the importance of the fragmentation functions, it is
unfortunate that uncertainties of the fragmentation functions have
not been estimated yet. The uncertainties have been investigated
extensively for unpolarized PDFs \cite{unpol-error}, polarized PDFs
\cite{pol-error}, and nuclear PDFs \cite{npdf-error}.
In order to extract any information from the hadron-production processes,
reliability regions of the fragmentation functions should be shown in
comparison with actual data. In particular, it is known that there are
large differences between the functions of KKP and Kretzer, for example
between their gluon functions. An error analysis should be done for
the fragmentation functions.

In this paper, we analyze the $e^+e^-$ data for obtaining the fragmentation
functions and their uncertainties in both LO and NLO. In particular, this
analysis is the first attempt for showing the uncertainties
of the fragmentation functions . Therefore, our works focus on
\begin{itemize}
\vspace{-0.15cm}
\item determination of the fragmentation functions
      and their uncertainties for pion, kaon, and proton
      in both LO and NLO,
\vspace{-0.15cm}
\item roles of NLO terms on the fragmentation-function determination,
      namely on their uncertainties,
\item comparison with other analysis results by considering
      the uncertainties.
\end{itemize}
\vspace{-0.15cm}
The functions are defined at an initial $Q^2$ point in terms of
a number of parameters, which are determined so as to explain
the $e^++e^- \rightarrow h +X$ data, where $h=\pi^\pm$, $K^\pm$,
and $p/\bar p$. The uncertainties are calculated by the Hessian method,
which has been used for obtaining the PDF uncertainties in
Refs. \cite{unpol-error, pol-error,npdf-error}.

This paper is organized as follows. 
A formalism is discussed in Sec. \ref{formalism} for hadron-production
cross sections in the $e^+e^-$ annihilation with the fragmentation functions.
In Sec. \ref{analysis}, our analysis method is described for determining
the fragmentation functions. Analysis results are explained in
Sec. \ref{results}. The results are summarized in Sec. \ref{summary}.
In Appendix \ref{other-hadrons}, we explain how to use the obtained
functions to other hadrons, $\pi^0$, $K^0$, $\bar K^0$, $n$, and $\bar n$.

\section{Formalism}
\label{formalism}

\subsection{Hadron production in $e^+ e^-$ annihilations}
\label{e-e+}

The cross section for the reaction $e^+ +e^- \rightarrow h+X$ is described
by two step processes \cite{hm-book}. The first part is to create
a quark-antiquark pair by the reaction $e^+ e^- \rightarrow q\bar q$,
and higher-order corrections such as $e^+ e^- \rightarrow q\bar q g$
are also taken into account in the NLO analysis. The second part is to
create a hadron $h$ from quark ($q$), antiquark ($\bar q$), or gluon ($g$),
and this process is called fragmentation.

The fragmentation function is defined by the hadron-production
cross section and the total hadronic cross section
\cite{esw-book}:
\begin{equation}  
F^h(z,Q^2) = \frac{1}{\sigma_{tot}} 
\frac{d\sigma (e^+e^- \rightarrow hX)}{dz} ,
\label{eqn:def-ff}
\end{equation}
where $Q^2$ is the virtual photon or $Z$ momentum squared
in $e^+e^- \rightarrow \gamma, Z$
and it is given by $Q^2=s$ with the center-of-mass energy $\sqrt{s}$.
The variable $z$ is defined by the energy fraction: 
\begin{equation}   
z \equiv \frac{E_h}{\sqrt{s}/2} = \frac{2E_h}{Q},
\label{eqn:def-z}
\end{equation}
where $E_h$ and $\sqrt{s}/2$  are the hadron and beam energies,
respectively. Namely, it indicates the hadron energy scaled 
to the beam energy.

The total cross section is described by the $q\bar q$-pair creation
processes, $e^+e^- \rightarrow \gamma \rightarrow q\bar q$ and
$e^+e^- \rightarrow Z \rightarrow q\bar q$, with higher-order
corrections:
\begin{equation}   
\sigma_{tot}=\sum_q \sigma_0^{q} (s)
   \left [ 1 + \frac{\alpha_s(Q^2)}{\pi} \right ] ,
\end{equation}
where the perturbative correction is included up to the NLO.
In the LO analysis, the second term is removed.
The electroweak cross section for producing a $q\bar q$ pair is given by
\cite{qqbar-cross}
\begin{align}
\sigma_0^{q} (s) = \frac{4 \pi \alpha^2}{s} \, 
  & [ \, e_q^2+2 e_q c_V^e c_V^q \rho_1(s) 
\nonumber \\
  &  + (c_V^{e\, 2}+c_A^{e\, 2})(c_V^{q\, 2}+c_A^{q\, 2}) \rho_2(s) \, ],
\end{align}
where the terms with $\rho_1(s)$ and $\rho_2(s)$ come from $\gamma$-$Z$
interference and $Z$ processes, and they are given by
\begin{align}
\rho_1(s) = & \frac{1}{4\sin^2\theta_W\cos^2\theta_W} 
       \frac{s(M_Z^2-s)}{(M_Z^2-s)^2+M_Z^2\Gamma_Z^2}, 
\nonumber \\
\rho_2(s) = & \left(\frac{1}{4\sin^2\theta_W\cos^2\theta_W}\right)^2
       \frac{s^2}{(M_Z^2-s)^2+M_Z^2\Gamma_Z^2}. 
\end{align}
Here, $\alpha$ is the fine structure constant in Quantum Electrodynamics
(QED), $e_q$ is a quark charge, and $\theta_W$ is the weak-mixing angle. 
The mass and width of $Z$ are denoted by $M_Z$ and $\Gamma_Z$, respectively.
The vector and axial-vector couplings $c_V^f$ and $c_A^f$ of a fermion $f$
are expressed by the third component of the weak isospin $T_f^3$ and
the fermion charge $e_f$:
\begin{equation}
c_V^f = T_f^3 -2 e_f \sin^2 \theta_W, \ \ \ 
c_A^f = T_f^3 .
\end{equation}
Actual expressions in terms of $\sin^2 \theta_W$ are
$c_V^e=-\frac{1}{2}+2\sin^2 \theta_W$ and $c_A^e=-\frac{1}{2}$
for the electron, 
$c_V^u=+\frac{1}{2}-\frac{4}{3}\sin^2 \theta_W$ and $c_A^u=+\frac{1}{2}$
for up, charm, and top quarks, and
$c_V^d=-\frac{1}{2}+\frac{2}{3}\sin^2 \theta_W$ and $c_A^d=-\frac{1}{2}$
for down, strange, and bottom quarks \cite{hm-book}.

\subsection{Fragmentation functions and their $Q^2$ evolution}
\label{ffs-q2}

The fragmentation process occurs from primary quarks, antiquarks,
and gluons, so that $F^h(z,Q^2)$ is expressed by the sum of
their contributions \cite{esw-book}:
\begin{equation}  
F^h(z,Q^2) = \sum_i C_i(z,\alpha_s) \otimes D_i^h (z,Q^2),
\label{eqn:def-ffqqbarg}
\end{equation}
where $D_i^h(z,Q^2)$ is a fragmentation function of the hadron $h$ 
from a parton $i$ ($=u,\ d,\ s,\ \cdot\cdot\cdot,\ g$),
$C_i(z,\alpha_s)$ is a coefficient function, and
the convolution integral $\otimes$ is defined by
\begin{equation}
f (z) \otimes g (z) = \int^{1}_{z} \frac{dy}{y}
            f(y) g\left(\frac{z}{y} \right)  .
\end{equation}
The function $D_i^h(z,Q^2)$ indicates the probability to find
the hadron $h$ from a parton $i$ with the energy fraction $z$.
The coefficient functions are calculated in perturbative QCD,
and the NLO results are listed, for example, in
Refs. \cite{kkp,kretzer,qqbar-cross} for the modified minimal
subtraction ($\overline {\rm MS}$) scheme. The cross section is
split into longitudinal and transverse components for
the virtual $\gamma$ or $Z$, so that the coefficient functions
are also expressed by these components.

$Q^2$ evolution for the fragmentation functions is calculated
in the same way as the one for the PDFs by using the timelike DGLAP
(Dokshitzer-Gribov-Lipatov-Altarelli-Parisi) evolution equations.
For example, the flavor-singlet evolution is given by \cite{esw-book}
\begin{align}
\frac{\partial}{\partial \ln Q^2}
\left(\begin{array}{c}
  D_S^h (z,Q^2) \\
  D_g^h (z,Q^2)
\end{array} \right) = & \frac{\alpha_s (Q^2)}{2\pi} \,
\left( \begin{array}{cc}
  P_{qq}(z) & 2 N_f P_{gq}(z) \\
  P_{qg}(z) & P_{gg}(z) \\
\end{array} \right) 
\nonumber \\
& \otimes
\left( \begin{array}{c}
  D_S^h (z,Q^2) \\
  D_g^h (z,Q^2)
\end{array} \right) ,
\label{eqn:evolution}
\end{align}
where $D_S^h (z,Q^2)$ denotes the singlet function
$D_S^h (z,Q^2)=\sum_q [ D_q^h (z,Q^2) +  D_{\bar q}^h (z,Q^2) ]$,
and $N_f$ is the number of quark flavors.
One should be careful that the off-diagonal elements
$P_{gq}(z)$ and $P_{qg}(z)$ are interchanged in the splitting-function
matrix from the PDF case. The splitting functions are the same
in the LO evolution of the PDFs; however, they are different in the NLO.
Explicit forms of the splitting functions are provided in
Refs. \cite{esw-book,splitting}.

The evolution equations are essentially the same as the PDF case,
so that the same numerical method can be applied for obtaining 
a solution. The equations are solved by direct integrations 
in the $z$ space as explained in Ref. \cite{evolution}.
A slightly modified numerical approach with the Simpson's integration
method is used in this analysis, and its evolution results are
independently checked by using a Gauss-Legendre quadrature.
Actual evolutions are calculated for the functions
$D_{q^\pm}^h = D_{q}^h \pm D_{\bar q}^h$, $D_S^h$, and $D_g^h$. 
Then, the quark
and antiquark functions are obtained by their combinations:
$D_{q}^h = (D_{q^+}^h +D_{q^-}^h)/2$ and
$D_{\bar q}^h = (D_{q^+}^h -D_{q^-}^h)/2$.

\section{\label{analysis} Analysis method}

\subsection{\label{paramet} Parametrization}

The fragmentation functions are expressed in terms of a number of
parameters at the initial $Q^2$ ($\equiv Q_0^2$) in the same way
as the PDF analysis \cite{unpol-error,pol-error,npdf-error}.
Since they should vanish at $z$=1, a simple polynomial form 
is taken:
\begin{equation}
D_i^h(z,Q_0^2) = N_i^h z^{\alpha_i^h} (1-z)^{\beta_i^h} ,
\end{equation}
where $N_i^h$, $\alpha_i^h$, and $\beta_i^h$ are parameters to be
determined by a $\chi^2$ analysis of $e^+e^- \rightarrow hX$ data.
This kind of polynomial form has been assumed in the parametrization
studies of the fragmentation functions \cite{kkp,kretzer}. 
The scale $Q_0^2$ is taken at mass thresholds $m_c^2$ and $m_b^2$
for charm and bottom functions, where $m_c$ and $m_b$ are charm-
and bottom-quark masses, respectively.
For the hadron $h$, we have pions ($\pi^+ +\pi^-$), kaons ($K^+ +K^-$),
and proton/anti-proton ($p+\bar p$). 

From actual $\chi^2$ analysis trials, we found that it is more appropriate
to take the second moment $M_i^h$ as a parameter instead of $N_i^h$.
This is because there exists an energy sum rule
\begin{equation}
\sum_h M_i^h = \sum_h \int_0^1 dz \, z \, D_i^h (z,Q^2) = 1 ,
\label{eqn:sum}
\end{equation}
for the function $D_i^h (z,Q^2)$, so that $M_i^h$ should not exceed one:
\begin{equation}
M_i^h < 1 .
\label{eqn:gamma_i}
\end{equation}
This constraint should be imposed in a global analysis. Another advantage
is that the physical meaning of $M_i^h$ is clear. It is the energy
fraction for the hadron $h$ which is created from the parton $i$.
These parameters $N_i^h$ and $M_i^h$ are related with each other by
\begin{equation}
N_i^h = \frac{M_i^h}{B(\alpha_i^h+2, \beta_i^h+1)},
\label{eqn:nih}
\end{equation}
where $B(\alpha_i^h+2, \beta_i^h+1)$ is the beta function.

A general principle of our parametrization is to use
a common function for favored fragmentation functions
from up and down quarks. A separate function is used
for a favored one from a strange quark by considering
the mass difference. Different functions are assigned for disfavored ones.
Here, the favored means the fragmentation from a quark which exits 
in the hadron $h$ as a constituent in the naive SU(6) quark model.
The disfavored means the fragmentation from a sea quark.
A flavor symmetric form is assumed for the fragmentation
functions from light sea-quarks (up, down, and strange sea-quarks).
Although it is known that light sea-quark distributions are not
flavor symmetric in the unpolarized PDFs \cite{flavor3},
there is no data to distinguish among the fragmentation
functions from different light sea-quarks.

The actual parametrization forms are shown for the pion,
kaon, and proton in the following.

\vspace{0.2cm}
\noindent
(1) Pion ($\pi^+$) 

Considering the constituent quark composition $\pi^+ (u \bar d)$,
we take the same favored fragmentation functions for $\pi^+$ from
$u$ and $\bar d$ quarks:
\begin{equation}
D_{u}^{\pi^+} (z,Q_0^2)  = D_{\bar d}^{\pi^+} (z,Q_0^2)
             = N_{u}^{\pi^+} z^{\alpha_{u}^{\pi^+}}
                (1-z)^{\beta_{u}^{\pi^+}}.
\end{equation}
The pion productions from $\bar u$, $d$, $s$, and $\bar s$ are
disfavored processes, and they are considered the same
at the initial scale: 
\begin{align}
D_{\bar u}^{\pi^+} (z,Q_0^2) & = D_{d}^{\pi^+} (z,Q_0^2)
\nonumber \\
& = D_{s}^{\pi^+} (z,Q_0^2)
  = D_{\bar s}^{\pi^+} (z,Q_0^2)
\nonumber \\
    & = N_{\bar u}^{\pi^+} z^{\alpha_{\bar u}^{\pi^+}}
         (1-z)^{\beta_{\bar u}^{\pi^+}} .
\end{align}
In addition, a fragmentation function from gluon is given by
\begin{equation}
D_{g}^{\pi^+} (z,Q_0^2) 
 = N_{g}^{\pi^+} z^{\alpha_{g}^{\pi^+}} (1-z)^{\beta_{g}^{\pi^+}} .
\end{equation}
These functions are provided at the initial scale $Q_0^2$ with
the parameters.

Different functions are assigned for productions
from heavy quarks because of mass differences:
\begin{align}
D_{c}^{\pi^+} (z,m_c^2) & = D_{\bar c}^{\pi^+} (z,m_c^2)
 = N_{c}^{\pi^+} z^{\alpha_{c}^{\pi^+}} (1-z)^{\beta_{c}^{\pi^+}} ,
\\
D_{b}^{\pi^+} (z,m_b^2) & = D_{\bar b}^{\pi^+} (z,m_b^2)
 = N_{b}^{\pi^+} z^{\alpha_{b}^{\pi^+}} (1-z)^{\beta_{b}^{\pi^+}} .
\end{align}
Thresholds for heavy quarks are $Q^2=m_c^2$ and $m_b^2$
in calculating $Q^2$ evolutions and the running coupling constant
$\alpha_s (Q^2)$.
However, the thresholds for the cross section are taken
$Q^2=4 m_c^2$ and $4 m_b^2$ \cite{kkp,kretzer}.

At different $Q^2$ ($\ne Q_0^2$), one should note that 
some equal relations do not hold in the NLO because of
a $q \rightarrow \bar q$ or $\bar q \rightarrow q$ splitting
\cite{flavor3}:
\begin{align}
D_{\bar u}^{\pi^+} (z,Q^2) & = D_{d}^{\pi^+} (z,Q^2)
     \ne D_{s}^{\pi^+} (z,Q^2)  ,
\end{align}
although $D_{u}^{\pi^+} (z,Q^2) = D_{\bar d}^{\pi^+} (z,Q^2)$,
$D_{s}^{\pi^+} (z,Q^2) = D_{\bar s}^{\pi^+} (z,Q^2)$,
$D_{c}^{\pi^+} (z,Q^2) = D_{\bar c}^{\pi^+} (z,Q^2)$, and
$D_{b}^{\pi^+} (z,Q^2) = D_{\bar b}^{\pi^+} (z,Q^2)$ are
still valid.

\vspace{0.3cm} \noindent
(2) Kaon ($K^+$) 

Parameters are assigned to $K^+$ fragmentation functions
in the same way by considering the constituent quark composition
$K^+ (u \bar s)$. The only difference is that the anti-strange
function is taken in a different form from the up-quark one:
\begin{alignat}{1}
D_{u}^{K^+} (z,Q_0^2) 
& = N_{u}^{K^+} z^{\alpha_{u}^{K^+}} (1-z)^{\beta_{u}^{K^+}},
\\
D_{\bar s}^{K^+} (z,Q_0^2)
& = N_{\bar s}^{K^+} z^{\alpha_{\bar s}^{K^+}} (1-z)^{\beta_{\bar s}^{K^+}},
\\
D_{\bar u}^{K^+} (z,Q_0^2) & = D_{d}^{K^+} (z,Q_0^2)
     = D_{\bar d}^{K^+} (z,Q_0^2)
\nonumber \\
& \! \! \! \! \! \! \! \! \! \! \! 
    = D_{s}^{K^+} (z,Q_0^2)
    = N_{\bar u}^{K^+} z^{\alpha_{\bar u}^{K^+}} 
      (1-z)^{\beta_{\bar u}^{K^+}}, 
\\
D_{c}^{K^+} (z,m_c^2) & = D_{\bar c}^{K^+} (z,m_c^2)
 = N_{c}^{K^+} z^{\alpha_{c}^{K^+}} (1-z)^{\beta_{c}^{K^+}} ,
\\
D_{b}^{K^+} (z,m_b^2) & = D_{\bar b}^{K^+} (z,m_b^2)
 = N_{b}^{K^+} z^{\alpha_{b}^{K^+}} (1-z)^{\beta_{b}^{K^+}} ,
\\
D_{g}^{K^+} (z,Q_0^2) 
 & = N_{g}^{K^+} z^{\alpha_{g}^{K^+}} (1-z)^{\beta_{g}^{K^+}} .
\end{alignat}
At general $Q^2$ ($\ne Q_0^2$), there are differences in the NLO
as explained in the pion case:
\begin{equation}
D_{\bar u}^{K^+} (z,Q^2) \ne D_{d}^{K^+} (z,Q^2)
                         \ne D_{s}^{K^+} (z,Q^2) .
\end{equation}
The relations 
$D_{d}^{K^+} (z,Q^2) = D_{\bar d}^{K^+} (z,Q^2)$,
$D_{c}^{K^+} (z,Q^2) = D_{\bar c}^{K^+} (z,Q^2)$, and
$D_{b}^{K^+} (z,Q^2) = D_{\bar b}^{K^+} (z,Q^2)$ are
still valid in the $Q^2$ evolution.

\vfill\eject
\vspace{0.3cm}
\noindent
(3) Proton ($p$) 

Proton fragmentation functions are also parametrized in the same way
by considering the constituent quark composition $p \, (uud)$:
\begin{alignat}{1}
D_{u}^{p} (z,Q_0^2) & = 2 \, D_{d}^{p} (z,Q_0^2) 
 = N_{u}^{p} z^{\alpha_{u}^{p}} (1-z)^{\beta_{u}^{p}}, 
\\
D_{\bar u}^{p} (z,Q_0^2) & = D_{\bar d}^{p} (z,Q_0^2)
= D_{s}^{p} (z,Q_0^2)
\nonumber \\
& \! \! \! \! \! \! \! \! 
    = D_{\bar s}^{p} (z,Q_0^2)
    = N_{\bar u}^{p} z^{\alpha_{\bar u}^{p}} (1-z)^{\beta_{\bar u}^{p}}, 
\\
D_{c}^{p} (z,m_c^2) & = D_{\bar c}^{p} (z,m_c^2)
 = N_{c}^{p} z^{\alpha_{c}^{p}} (1-z)^{\beta_{c}^{p}} ,
\\
D_{b}^{p} (z,m_b^2) & = D_{\bar b}^{p} (z,m_b^2)
 = N_{b}^{p} z^{\alpha_{b}^{p}} (1-z)^{\beta_{b}^{p}} ,
\\
D_{g}^{p} (z,Q_0^2) 
 & = N_{g}^{p} z^{\alpha_{g}^{p}} (1-z)^{\beta_{g}^{p}} .
\end{alignat}
The major difference from the mesons is the factor of two
in $D_{u}^{p} (z,Q_0^2) = 2 \, D_{d}^{p} (z,Q_0^2)$ 
which is suggested simply by considering valence-quark structure
with a flavor symmetry \cite{kkp}. In order to produce
a baryon from a quark, two $q\bar q$ pairs need to be created.
If the initial quark is up quark, two creation process,
$(u\bar u)(d \bar d)$ and $(d\bar d)(u \bar u)$,
should contribute to the proton formation, whereas only
the $(u\bar u)(u \bar u)$ process contributes if the down quark is
in the initial state. It leads to the factor of two although
it is a naive counting estimate. This factor is also the same
in a spectator di-quark model in the flavor symmetric case
\cite{diquark}.

At different $Q^2$, one should be careful about the following
inequalities in the NLO:
\begin{alignat}{1}
D_{u}^{p} (z,Q^2) & \ne 2 \, D_{d}^{p} (z,Q^2) ,
\\
D_{\bar u}^{p} (z,Q^2) & \ne D_{\bar d}^{p} (z,Q^2)
                         \ne D_{s}^{p} (z,Q^2) .
\end{alignat}
The equalities for strange, charm, and bottom quarks,
$D_{s}^{p} (z,Q^2) = D_{\bar s}^{p} (z,Q^2)$,
$D_{c}^{p} (z,Q^2) = D_{\bar c}^{p} (z,Q^2)$, and
$D_{b}^{p} (z,Q^2) = D_{\bar b}^{p} (z,Q^2)$, are
not changed by the $Q^2$ evolution.

\vspace{0.3cm} \noindent
(4) $\pi^-$, $K^-$, and $\bar p$

In the constituent quark model, $\pi^-$, $K^-$, and $\bar p$ are expressed
$\pi^- (\bar u d)$, $K^-(\bar u s)$, and $\bar p \, (\bar u\bar u\bar d)$.
Using the charge symmetry, we relate the fragmentation functions
of $\pi^-$ to the ones of $\pi^+$:
\begin{equation}
D_{q}^{\pi^-, \, K^-, \, \bar p} (z,Q^2) 
= D_{\bar q}^{\pi^+, \, K^+, \, p} (z,Q^2) .
\label{eqn:pi-k-pbar-q}
\end{equation}
The gluonic functions are the same:
\begin{equation}
D_{g}^{\pi^-, \, K^-, \, \bar p} (z,Q^2)
= D_{g}^{\pi^+, \, K^+, \, p} (z,Q^2)  .
\label{eqn:pi-k-pbar-g}
\end{equation}
The functions for $\pi^0$, $K^0$, $\bar K^0$, $n$, and $\bar n$ are
also obtained by using the charge symmetry, and it is explained in
Appendix \ref{other-hadrons}.

After all, the following parameters are used for the pion
kaon, and proton in our global analysis:
\begin{align}
\text{pion:} \ & 
   \alpha_{u}^{\pi^+} \! \! , \, \beta_{u}^{\pi^+} \! \! , \, 
               M_{u}^{\pi^+} \! \! , \,
   \alpha_{\bar u}^{\pi^+} \! \! , \, \beta_{\bar u}^{\pi^+} \! \! , \, 
               M_{\bar u}^{\pi^+} \! \! , \,
\nonumber \\
 & 
   \alpha_{c}^{\pi^+} \! \! , \, \beta_{c}^{\pi^+} \! \! , \, 
               M_{c}^{\pi^+} \! \! ,
   \alpha_{b}^{\pi^+} \! \! , \, \beta_{b}^{\pi^+} \! \! , \, 
               M_{b}^{\pi^+} \! \! , \,
   \alpha_{g}^{\pi^+} \! \! , \, \beta_{g}^{\pi^+} \! \! , \, 
               M_{g}^{\pi^+} \! \! , 
\nonumber \\
\text{kaon:} \ & 
   \alpha_{u}^{K^+} \! \! , \, \beta_{u}^{K^+} \! \! , \, 
           M_{u}^{K^+} \! \!, \,
   \alpha_{\bar s}^{K^+} \! \! , \, \beta_{\bar s}^{K^+} \! \! , \, 
           M_{\bar s}^{K^+} \! \! , \,
   \alpha_{\bar u}^{K^+} \! \! , \, \beta_{\bar u}^{K^+} \! \! , \, 
           M_{\bar u}^{K^+} \! \! , \,
\nonumber \\
 &   
   \alpha_{c}^{K^+} \! \! ,  \, \beta_{c}^{K^+} \! \! , \, 
           M_{c}^{K^+} \! \! , \,
   \alpha_{b}^{K^+} \! \! , \, \beta_{b}^{K^+} \! \! , \,  
           M_{b}^{K^+} \! \! , \, 
   \alpha_{g}^{K^+} \! \! , \, \beta_{g}^{K^+} \! \! , \, 
           M_{g}^{K^+} \! \! ,      
\nonumber \\
\text{proton:} \ & 
   \alpha_{u}^{p}, \, \beta_{u}^{p}, \, M_{u}^{p}, \,
   \alpha_{\bar u}^{p}, \, \beta_{\bar u}^{p}, \, M_{\bar u}^{p}, \,
\nonumber \\
 &   
   \alpha_{c}^{p}, \, \beta_{c}^{p}, \, M_{c}^{p}, \
   \alpha_{b}^{p}, \, \beta_{b}^{p}, \, M_{b}^{p}, \, 
   \alpha_{g}^{p}, \, \beta_{g}^{p}, \, M_{g}^{p}.
\nonumber
\end{align}

\subsection{Experimental data}
\label{data} 

The fragmentation functions are determined by the charged-hadron production
data of $e^+ +e^- \rightarrow h^\pm +X$. Used data are those from
the measurements of TASSO \cite{tasso12_30,tasso14_22,tasso34_44},
TPC \cite{tpc29}, HRS \cite{hrs29}, TOPAZ \cite{topaz58}, SLD \cite{sld91},
ALEPH \cite{aleph91}, OPAL \cite{opal91}, and 
DELPHI \cite{delphi91,delphi91-2}.
There are data from MARK-II \cite{mark-II29} and JADE \cite{jade34}
collaborations; however, they are not used in our analysis because
of the kinematical condition $z>0.1$, which is explained in the following.
References, center-of-mass (c.m.) energies, and numbers of the data are
listed for used data sets in Tables \ref{tab:exp-pion}, \ref{tab:exp-kaon},
and \ref{tab:exp-proton} \cite{durham}.
The ALEPH, DELPHI, and OPAL data are taken
at CERN (European Organization for Nuclear Research),
TASSO at DESY (German Electron Synchrotron),
TOPAZ at KEK (High Energy Accelerator Research Organization), and
HRS, SLD, and TPC at SLAC (Stanford Linear Accelerator Center). 
In the DELPHI and SLD measurements, the light-quark and heavy-quark events
are separated. 

Since the perturbative QCD is applied in the $Q^2$ evolution
calculations, the data with $Q^2>1$ GeV$^2$ are used in our analysis.
Furthermore, small-$z$ data are excluded because of soft-gluon emission.
The minimum $z$ value is $z_{min}=0.1$ for the data at $\sqrt{s}<M_Z$
and $z_{min}=0.05$ for the data at $\sqrt{s}=M_Z$.
Resummation effects of soft-gluon logarithms need to be clarified
in order to include the small-$z$ data into the analysis \cite{resum}. 
We do not include three-jet and unidentified-hadron data \cite{kkp,bfgw01}.
There are ambiguities for extracting the gluon
fragmentation function from three-jet cross sections \cite{kretzer}.
In oder to describe the unidentified-hadron cross sections, all the relevant
hadrons need to be included in the analysis, whereas only the pions, kaons,
proton, and anti-proton are taken into account in our analysis.

\begin{table}[t]
\caption{Experiments, references, center-of-mass energies,
     and numbers of data points are listed for used data sets
     of $e^+ +e^- \rightarrow \pi^\pm +X$ \cite{durham}.}
\label{tab:exp-pion}
\begin{ruledtabular}
\begin{tabular*}{\hsize}
{l@{\extracolsep{0ptplus1fil}}c@{\extracolsep{0ptplus1fil}}c
@{\extracolsep{0ptplus1fil}}c}
experiment & ref.    & $\sqrt{s}$ & \# of data \\
\colrule\colrule
                     &                 &                         &        \\
TASSO                & \cite{tasso12_30,tasso14_22,tasso34_44}
                                       & 12,14,22,30,34,44       &  29    \\
TPC                  & \cite{tpc29}    & 29                      &  18    \\
HRS                  & \cite{hrs29}    & 29                      & \, 2   \\
TOPAZ                & \cite{topaz58}  & 58                      & \, 4   \\
SLD                  & \cite{sld91}    & 91.28                   &  29    \\
SLD (u,d,s quark)    & \cite{sld91}    & 91.28                   &  29    \\
SLD (c quark)        & \cite{sld91}    & 91.28                   &  29    \\
SLD (b quark)        & \cite{sld91}    & 91.28                   &  29    \\
ALEPH                & \cite{aleph91}  & 91.2  \,                &  22    \\
OPAL                 & \cite{opal91}   & 91.2  \,                &  22    \\
DELPHI               & \cite{delphi91} & 91.2  \,                &  17    \\
DELPHI (u,d,s quark) & \cite{delphi91} & 91.2  \,                &  17    \\
DELPHI (b quark)     & \cite{delphi91} & 91.2  \,                &  17    \\
\colrule
total                &                 &                         & 264 \, \\
\end{tabular*}
\end{ruledtabular}
\end{table}

\begin{table}[h]
\vspace{-0.35cm}
\caption{Experimental information is listed for the used data
     of $e^+ +e^- \rightarrow K^\pm +X$ \cite{durham}.}
\label{tab:exp-kaon}
\begin{ruledtabular}
\begin{tabular*}{\hsize}
{l@{\extracolsep{0ptplus1fil}}c@{\extracolsep{0ptplus1fil}}c
@{\extracolsep{0ptplus1fil}}c}
experiment & ref.    & $\sqrt{s}$ & \# of data \\
\colrule\colrule
                     &                 &                         &        \\
TASSO                & \cite{tasso12_30,tasso14_22,tasso34_44}
                                       & 12,14,22,30,34          &  18    \\
TPC                  & \cite{tpc29}    & 29                      &  17    \\
HRS                  & \cite{hrs29}    & 29                      & \, 3   \\
TOPAZ                & \cite{topaz58}  & 58                      & \, 3   \\
SLD                  & \cite{sld91}    & 91.28                   &  29    \\
SLD (u,d,s quark)    & \cite{sld91}    & 91.28                   &  29    \\
SLD (c quark)        & \cite{sld91}    & 91.28                   &  29    \\
SLD (b quark)        & \cite{sld91}    & 91.28                   &  28    \\
ALEPH                & \cite{aleph91}  & 91.2  \,                &  18    \\
OPAL                 & \cite{opal91}   & 91.2  \,                &  10    \\
DELPHI               & \cite{delphi91,delphi91-2}
                                       & 91.2  \,                &  27    \\
DELPHI (u,d,s quark) & \cite{delphi91} & 91.2  \,                &  17    \\
DELPHI (b quark)     & \cite{delphi91} & 91.2  \,                &  17    \\
\colrule
total                &                 &                         & 245 \, \\
\end{tabular*}
\end{ruledtabular}
\end{table}

\begin{table}[h!]
\vspace{-0.35cm}
\caption{Experimental information is listed for the used data
     of $e^+ +e^- \rightarrow p/\bar p +X$ \cite{durham}.}
\label{tab:exp-proton}
\begin{ruledtabular}
\begin{tabular*}{\hsize}
{l@{\extracolsep{0ptplus1fil}}c@{\extracolsep{0ptplus1fil}}c
@{\extracolsep{0ptplus1fil}}c}
experiment & ref.    & $\sqrt{s}$ & \# of data \\
\colrule\colrule
                     &                 &                         &        \\
TASSO                & \cite{tasso12_30,tasso14_22,tasso34_44}
                                       & 12,14,22,30,34          &  24    \\
TPC                  & \cite{tpc29}    & 29                      &  17    \\
HRS                  & \cite{hrs29}    & 29                      & \, 4   \\
TOPAZ                & \cite{topaz58}  & 58                      & \, 3   \\
SLD                  & \cite{sld91}    & 91.28                   &  30    \\
SLD (u,d,s quark)    & \cite{sld91}    & 91.28                   &  30    \\
SLD (c quark)        & \cite{sld91}    & 91.28                   &  30    \\
SLD (b quark)        & \cite{sld91}    & 91.28                   &  26    \\
ALEPH                & \cite{aleph91}  & 91.2  \,                &  18    \\
OPAL                 & \cite{opal91}   & 91.2  \,                &  10    \\
DELPHI               & \cite{delphi91,delphi91-2}
                                       & 91.2  \,                &  23    \\
DELPHI (u,d,s quark) & \cite{delphi91} & 91.2  \,                &  17    \\
DELPHI (b quark)     & \cite{delphi91} & 91.2  \,                &  17    \\
\colrule
total                &                 &                         & 249 \, \\
\end{tabular*}
\end{ruledtabular}
\end{table}

\begin{figure}[t]
\includegraphics[width=0.40\textwidth]{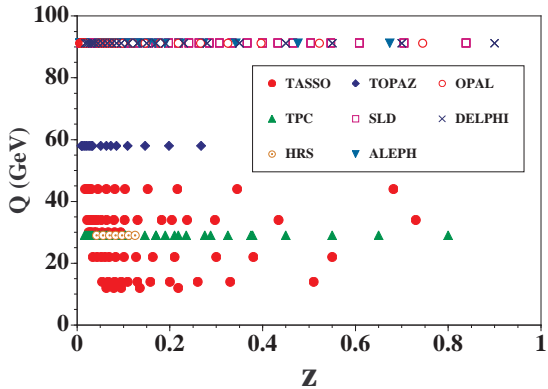}
\vspace{-0.3cm}
\caption{(Color online) Kinematical range is shown by
         $z$ and $Q (=\sqrt{s})$ values for
         the pion data.}
\label{fig:zq2}
\end{figure}

The kinematical range is shown in Fig. \ref{fig:zq2} by the variables
$z$ and $Q (\equiv\sqrt{Q^2}=\sqrt{s})$ for the pion data. Here, the data
are also shown in the small $z$ region ($z<0.1$ or $z<0.05$), where
they are not included in the actual analysis. The range of $Q$ is wide
from the low-energy TASSO data to high-energy SLD, ALEPH, OPAL, and 
DELPHI data at $Q=M_Z$. Many data are collected in the small-$z$ region
($z <0.4$), whereas the large-$z$ data are scarce because cross sections
are very small. 
 
\subsection{$\chi^2$ analysis}
\label{chi2analysis} 

The total $\chi^2$ is calculated in comparison with the data for
the fragmentation function $F^h(z,Q^2)$ of the hadron $h$ in
Eq. (\ref{eqn:def-ff}). The scale $Q^2$ is equal to the c.m. energy
squared $s$ in Tables \ref{tab:exp-pion}, \ref{tab:exp-kaon}, and
\ref{tab:exp-proton}. The initial scale, where the fragmentation functions
for up, down, and strange quarks are defined, is taken $Q_0^2$=1 GeV$^2$.
Since the $e^+ e^-$ data are measured at large $Q^2$ as shown in the tables,
we could have taken a larger $Q_0^2$ in the range where perturbative QCD
can be applied undoubtedly without worrying about higher-twist and
higher-order corrections. However, the fragmentation functions are 
practically used also in a relatively small $Q^2$ region, for example,
in hadron-production process in deep inelastic lepton scattering
\cite{sidis} and proton-proton collisions \cite{rhic-pi}.
For applications of obtained functions to such processes, it is better
to supply a code for calculating the functions even at small $Q^2$
($\sim$ 1 GeV$^2$). Considering this situation, we decided to take
$Q_0^2$=1 GeV$^2$. The heavy-quark functions are defined at 
the mass thresholds. The quark masses are $m_c$=1.43 GeV and
$m_b$=4.3 GeV as given in Ref. \cite{th-mass}. 

The theoretical functions should be obtained at the same experimental
$Q^2$ points for calculating $\chi^2$. The $Q^2$ evolution
is calculated by the timelike DGLAP equations as explained in
Sec. \ref{ffs-q2}. Then, the evolved functions are convoluted 
with the coefficient functions in Eq. (\ref{eqn:def-ffqqbarg}).
These calculations depend on the choice of the scale parameter
$\Lambda$. It is taken $\Lambda^{(4)}_{LO}$=0.220 GeV or
$\Lambda^{(4)}_{NLO}$=0.323 GeV for four flavors \cite{th-mass}. 
The $\overline {\rm MS}$ scheme is used in the NLO. 
Below the charm threshold and above the bottom one, these values
are converted to corresponding ones for three and five flavors
so that the coupling constant $\alpha_s$ is continuous
at the thresholds.

The total $\chi^2$ is then calculated by
\begin{equation}
\chi^2 = \sum_j \frac{(F_{j}^{data}-F_{j}^{theo})^2}
                     {(\sigma_j^{data})^2},
\label{eqn:chi2}
\end{equation}
where $F_{j}^{data}$ and $F_{j}^{theo}$ are experimental and theoretical
values of $F^h (z,Q^2)$, respectively, at the same experimental 
$Q^2$ point. The experimental errors are calculated from systematic and
statistical errors by
$(\sigma_j^{data})^2 = (\sigma_j^{sys})^2 + (\sigma_j^{stat})^2$.
The assigned parameters are determined so as to obtain the minimum $\chi^2$.
The optimization of the functions is done by the CERN subroutine
{\tt MINUIT} \cite{minuit}.

\subsection{Uncertainties of fragmentation functions}
\label{uncertainties} 

Uncertainties of the PDFs were estimated in Refs. 
\cite{unpol-error,pol-error,npdf-error}. Two methods have been developed:
the Hessian and Lagrange-multiplier methods. Although technical details
are described in these references, outline of the Hessian method is 
explained because it is used in our analysis and uncertainties
have not been investigated in the fragmentations functions.

The parameters are denoted $\xi_i$ ($i$=1, 2, $\cdot \cdot \cdot$, $N$),
where $N$ is the number of the parameters. We expand $\chi^2$ around
the minimum $\chi^2$ point $\hat \xi$:
\begin{equation}
 \Delta \chi^2 (\xi) = \chi^2(\hat{\xi}+\delta \xi)-\chi^2(\hat{\xi})
        =\sum_{i,j} H_{ij}\delta \xi_i \delta \xi_j \ ,
\label{eq:chi2expand}
\end{equation}
where only the leading quadratic term is kept, and
the Hessian $H_{ij}$ is the second derivative matrix. 
The confidence region is given in the parameter space by
supplying a value of $\Delta \chi^2$.
It is known that the confidence level is 68\% for $\Delta \chi^2$=1
if the number of the parameters is one ($N=1$). The $\Delta \chi^2$
value needs to be changed in a general case of $N \ne 1$.
Assuming correspondence between the confidence level of
a normal distribution in multi-parameter space and
the one of a $\chi^2$ distribution with N degree of freedom,
we have the confidence level $P$:
\begin{equation}
        P=\int_0^{\Delta \chi^2} \frac{1}{2\ \Gamma(N/2)} 
        \left(\frac{S}{2}\right)^{\frac{N}{2}-1} 
               \exp\left(-\frac{S}{2} \right) dS \ ,
\label{eq:dchi2}
\end{equation}
where $\Gamma(N/2)$ is the Gamma function. The value of $\Delta \chi^2$
is taken so that the confidence level becomes the one-$\sigma$-error 
range, namely $P=0.6826$ \cite{pol-error,npdf-error}.
The details of this $\Delta \chi^2$ choice are
explained in Ref. \cite{del-chi2}. The $\Delta \chi^2$ value
is numerically calculated by using Eq. (\ref{eq:dchi2}). 
For example, $\Delta \chi^2=15.94$ is obtained for $N=14$.
The Hessian matrix is obtained by running the subroutine {\tt MINUIT}.
From this Hessian, $\Delta \chi^2$, and derivatives of the fragmentation
functions with respect to the parameters, the uncertainties are
calculated by
\begin{equation}
[\delta D_i^h (z)]^2=\Delta \chi^2 \sum_{j,k}
\left( \frac{\partial D_i^h (z,\xi)}{\partial \xi_j}  \right)_{\hat\xi}
H_{jk}^{-1}
\left( \frac{\partial D_i^h (z,\xi)}{\partial \xi_k}  \right)_{\hat\xi}
\, .
\label{eqn:ddih}
\end{equation}

\vfill\eject
\section{\label{results} Results}

Analysis results are explained. In Sec. \ref{comp}, optimized parameters
are shown, and $\chi^2$ contributions from data sets are listed. 
Then, fit results are compared with experimental data. 
In Sec. \ref{ffs}, the obtained fragmentation functions and their
uncertainties are shown. They are compared with other parametrization
results in Sec. \ref{other-ffs}.

\subsection{Comparison with experimental data}
\label{comp}

Obtained parameters in the LO and NLO are listed in Tables
\ref{table:parameters-pion}, \ref{table:parameters-kaon},
and \ref{table:parameters-proton} for the pion, kaon, and proton,
respectively. In these analyses, it was very difficult to determine
the gluon functions, so that we decided to fix some parameters.
In trial analyses, we found that the gluon function
for the pion ($z D_g^{\pi^+}$) tends to be peaked at $z \sim 0.2$.
The parameter $\beta_g$ controls its functional behavior at large $z$.
Since the gluon function becomes very small at large $z$, the value of
$\beta_g$ does not affect the $\chi^2$ value to a significant amount
if the function is peaked at small $z$. Therefore, the parameter $\beta_g$
is fixed as $\beta_g$=8 in the pion analyses.
If this value is taken much more than eight, the energy sum rule in
Eq. (\ref{eqn:sum}) is badly violated in the gluon part.
If it much less than eight, it is violated in the up-quark part.
The value $\beta_g$=8 is chosen to compromise these two issues. 
In the analysis of Kretzer \cite{kretzer}, one of
the gluon parameters is also fixed because the second moment is
equal to the average of the moments of up- and down-quark
fragmentation functions. 

In the kaon analyses, the gluon function tends to be peaked
at large $z$. Therefore, $\alpha_g$ becomes the parameter
which does not affect the total $\chi^2$ instead
of $\beta_g$. It is chosen $\alpha_g=5$ so that the moments
for gluon and up quark do not become too large to affect
the sum rule to a significant extent. 

It was also very difficult to determine the gluon distribution for
the proton. It indicates that the data are not sensitive to
the gluon function at this stage even in the NLO analysis.
Therefore, the moment of the gluon function is fixed at
the average of the moments for the favored and disfavored
functions \cite{kretzer}: $M_g = [(M_u +M_d)/2+M_{\bar u}]/2$.
Furthermore, the function tends to be peaked at large $z$,
so that the parameter $\alpha_g$ is fixed at $\alpha_g=5$
in the same way as the kaon case. The average of the moments
is motived by the following consideration. The process for
producing a proton from an initial up or down quark should contain
two quark-pair productions ($u\bar u$ and $d\bar d$ pairs or
two $u\bar u$ pairs), so that it is proportional to $g^4$
with the strong coupling constant $g$.
In the same way, the process from an initial $\bar u$ quark
is proportional to $g^6$, and the one from an initial gluon
is to $g^5$. Therefore, the gluon moment is, roughly speaking,
given by the average of favored ($u$ and $d$) and disfavored ($\bar u$)
moments. Such averages are also taken in Ref. \cite{kretzer}.

\newcommand{\ph}{\phantom{$-$}}
\begin{table}[t]
\caption{\label{table:parameters-pion}
         Parameters determined for the pion.}
\begin{ruledtabular}
\begin{tabular}{lccc} 
function             &  
 $M$                 & $\alpha$                   & $\beta$               \\
\hline
(LO)                 &   
                     &                            &                       \\
$D_u^{\pi^+}$        & 
 0.546 $\pm$ 0.085   &  $-$1.100 $\pm$ 0.183      &  1.282 $\pm$ 0.140    \\
$D_{\bar u}^{\pi^+}$ &  
 0.250 $\pm$ 0.068   &  $-$0.500 $\pm$ 0.301      &  5.197 $\pm$ 0.576    \\
$D_c^{\pi^+}$        & 
 0.305 $\pm$ 0.046   &  $-$1.007 $\pm$ 0.123      &  3.918 $\pm$ 0.236    \\
$D_b^{\pi^+}$        & 
 0.302 $\pm$ 0.023   &  $-$1.176 $\pm$ 0.045      &  5.805 $\pm$ 0.188    \\
$D_g^{\pi^+}$        & 
 0.115 $\pm$ 0.111   & \ph 1.405 $\pm$ 0.897      &  8.0 (fixed)          \\
\hline
(NLO)                & 
                     &                            &                       \\
$D_u^{\pi^+}$        & 
 0.401 $\pm$ 0.052   &  $-$0.963 $\pm$ 0.177      &  1.370  $\pm$ 0.144   \\
$D_{\bar u}^{\pi^+}$ & 
 0.094 $\pm$ 0.029   & \ph 0.718 $\pm$ 0.466      &  6.266  $\pm$ 0.808   \\
$D_c^{\pi^+}$        &
 0.178 $\pm$ 0.018   &  $-$0.845 $\pm$ 0.108      &  3.868  $\pm$ 0.323   \\
$D_b^{\pi^+}$        & 
 0.236 $\pm$ 0.009   &  $-$1.219 $\pm$ 0.042      &  5.668  $\pm$ 0.219   \\
$D_g^{\pi^+}$        & 
 0.238 $\pm$ 0.029   & \ph 1.943 $\pm$ 0.399      &  8.0 (fixed)          \\
\end{tabular}
\end{ruledtabular}
\end{table}

\begin{table}[h]
\vspace{+0.8cm}
\caption{\label{table:parameters-kaon}
         Parameters determined for the kaon}
\begin{ruledtabular}
\begin{tabular}{lccc} 
function             &  
 $M$                 & $\alpha$                 & $\beta$                \\
\hline
(LO)                 & 
                     &                          &                        \\
$D_u^{K^+}$          & 
 0.0922 $\pm$ 0.0419 &  0.588 $\pm$ 1.605       &  1.632 $\pm$ 0.812     \\
$D_{\bar s}^{K^+}$   & 
 0.1651 $\pm$ 0.0962 &  2.190 $\pm$ 2.871       &  2.829 $\pm$ 1.143     \\
$D_{\bar u}^{K^+}$   &
 0.0638 $\pm$ 0.0363 &  0.565 $\pm$ 0.460       &  7.093 $\pm$ 3.383     \\
$D_c^{K^+}$          & 
 0.0919 $\pm$ 0.0055 &  0.230 $\pm$ 0.157       &  4.549 $\pm$ 0.388     \\
$D_b^{K^+}$          & 
 0.0696 $\pm$ 0.0027 &  0.017 $\pm$ 0.110       &  8.808 $\pm$ 0.534     \\
$D_g^{K^+}$          & 
 0.0319 $\pm$ 0.0147 &   5.0 (fixed)            &  0.247 $\pm$ 0.162     \\
\hline
(NLO)                & 
                     &                          &                        \\
$D_u^{K^+}$          & 
 0.0740 $\pm$ 0.0268 &  $-$0.630  $\pm$ 0.629   &  1.310 $\pm$ 0.772     \\
$D_{\bar s}^{K^+}$   & 
 0.0878 $\pm$ 0.0506 & \ph 2.000  $\pm$ 2.913   &  2.800 $\pm$ 1.313     \\
$D_{\bar u}^{K^+}$   & 
 0.0255 $\pm$ 0.0173 & \ph 1.012  $\pm$ 0.939   &  8.000 $\pm$ 3.715     \\
$D_c^{K^+}$          & 
 0.0583 $\pm$ 0.0052 & \ph 0.527  $\pm$ 0.228   &  5.866 $\pm$ 0.636     \\
$D_b^{K^+}$          & 
 0.0522 $\pm$ 0.0024 & \ph 0.247  $\pm$ 0.126   &  11.212 $\pm$ 0.721 \, \\
$D_g^{K^+}$          & 
 0.0705 $\pm$ 0.0099 & \ph 5.0 (fixed)          &  0.810 $\pm$ 0.239     \\
\end{tabular}
\end{ruledtabular}
\end{table}

\begin{table}[h!]
\vspace{+0.8cm}
\caption{\label{table:parameters-proton}
         Parameters determined for the proton.}
\begin{ruledtabular}
\begin{tabular}{lccc} 
function             & 
  $M$                & $\alpha$                 & $\beta$                \\
\hline
(LO)                 & 
                     &                          &                        \\
$D_u^{p}$            & 
 0.0839 $\pm$ 0.0125 &  $-$0.814 $\pm$ 0.200    &   1.628 $\pm$ 0.324    \\
$D_{\bar u}^{p}$     & 
 0.0158 $\pm$ 0.0058 & \ph 0.866 $\pm$ 0.677    &   5.078 $\pm$ 1.400    \\
$D_c^{p}$            & 
 0.0241 $\pm$ 0.0015 & \ph 0.683 $\pm$ 0.359    &   7.375 $\pm$ 1.250    \\
$D_b^{p}$            & 
 0.0180 $\pm$ 0.0006 & \ph 0.071 $\pm$ 0.178    &   8.802 $\pm$ 0.839    \\
$D_g^{p}$            & 
$\frac{(M_u+M_d)/2+M_{\bar u}}{2}$ 
                     & \ph 5.0 (fixed)          &   2.927 $\pm$ 1.117    \\
\hline
(NLO)                &                          & 
                     &                           \\
$D_u^{p}$            & 
 0.0732 $\pm$ 0.0113 &  $-$1.022 $\pm$ 0.219   &  1.434  $\pm$ 0.268    \\
$D_{\bar u}^{p}$     & 
 0.0084 $\pm$ 0.0057 & \ph 1.779 $\pm$ 1.422   &  4.763  $\pm$ 1.882    \\
$D_c^{p}$            & 
 0.0184 $\pm$ 0.0017 & \ph 0.407 $\pm$ 0.373   &  6.784  $\pm$ 1.555    \\
$D_b^{p}$            & 
 0.0155 $\pm$ 0.0007 &  $-$0.203 $\pm$ 0.165   &  8.209  $\pm$ 0.950    \\
$D_g^{p}$            &
$\frac{(M_u+M_d)/2+M_{\bar u}}{2}$ 
                     & \ph 5.0 (fixed)         &  4.900  $\pm$ 2.046    \\
\end{tabular}
\end{ruledtabular}
\end{table}

\begin{table}[t]
\caption{Each $\chi^2$ contribution in the pion analysis.}
\label{tab:chi2-pion}
\begin{ruledtabular}
\begin{tabular*}{\hsize}
{l@{\extracolsep{0ptplus1fil}}c@{\extracolsep{0ptplus1fil}}c
@{\extracolsep{0ptplus1fil}}c}
experiment           & \# of data  & $\chi^2$ (LO)  & $\chi^2$ (NLO) \\
\colrule\colrule
TASSO                &    29       &  52.1          &  51.9          \\
TPC                  &    18       &  33.5          &  27.3          \\
HRS                  &  \, 2       & \,  1.1        &  \, 2.0        \\
TOPAZ                &  \, 4       & \,  2.6        &  \, 2.6        \\
SLD (all)            &    29       &  11.3          &  10.6          \\
SLD (u,d,s)          &    29       &  46.0          &  36.4          \\
SLD (c)              &    29       &  24.4          &  26.1          \\
SLD (b)              &    29       &  71.2          &  66.4          \\
ALEPH                &    22       &  22.8          &  24.0          \\
OPAL                 &    22       &  45.4          &  45.8          \\
DELPHI (all)         &    17       &  48.3          &  48.6          \\
DELPHI (u,d,s)       &    17       &  29.6          &  31.1          \\
DELPHI (b)           &    17       &  64.9          &  60.8          \\
\colrule
total                &   264 \,    &  453.2 \,      &  433.5 \,      \\
(/d.o.f.)            &             &  (1.81)        &  (1.73)        \\
\end{tabular*}
\end{ruledtabular}
\end{table}

\begin{table}[h]
\caption{Each $\chi^2$ contribution in the kaon analysis.}
\label{tab:chi2-kaon}
\begin{ruledtabular}
\begin{tabular*}{\hsize}
{l@{\extracolsep{0ptplus1fil}}c@{\extracolsep{0ptplus1fil}}c
@{\extracolsep{0ptplus1fil}}c}
experiment           & \# of data  & $\chi^2$ (LO)  & $\chi^2$ (NLO) \\
\colrule\colrule
TASSO                &    18       &  26.8          &  25.0          \\
TPC                  &    17       &  15.2          &  15.2          \\
HRS                  &  \, 3       & \,  0.6        &  \, 0.4        \\
TOPAZ                &  \, 3       & \,  0.5        &  \, 0.8        \\
SLD (all)            &    29       &  14.9          &  12.3          \\
SLD (u,d,s)          &    29       &  58.8          &  57.2          \\
SLD (c)              &    29       &  33.6          &  32.4          \\
SLD (b)              &    28       & 127.7          &  88.7          \\
ALEPH                &    18       &  10.3          &  12.8          \\
OPAL                 &    10       &  10.6          &  11.5          \\
DELPHI (all)         &    27       &  14.8          &  15.2          \\
DELPHI (u,d,s)       &    17       &  22.5          &  22.1          \\
DELPHI (b)           &    17       &  11.7          &  11.7          \\
\colrule
total                &   245 \,    &  348.2 \,      &  305.1 \,      \\
(/d.o.f.)            &             &  (1.53)        &  (1.34)        \\
\end{tabular*}
\end{ruledtabular}
\end{table}

\begin{table}[h!]
\caption{Each $\chi^2$ contribution in the proton/anti-proton analysis.}
\label{tab:chi2-proton}
\begin{ruledtabular}
\begin{tabular*}{\hsize}
{l@{\extracolsep{0ptplus1fil}}c@{\extracolsep{0ptplus1fil}}c
@{\extracolsep{0ptplus1fil}}c}
experiment           & \# of data  & $\chi^2$ (LO)  & $\chi^2$ (NLO) \\
\colrule\colrule
TASSO                &    24       &   34.9         &  33.6          \\
TPC                  &    17       &   22.6         &  23.5          \\
HRS                  &  \, 4       &   19.5         &  16.3          \\
TOPAZ                &  \, 3       & \, 3.6         &  \, 3.2        \\
SLD (all)            &    30       &   13.0         &  12.6          \\
SLD (u,d,s)          &    30       &   62.9         &  56.6          \\
SLD (c)              &    30       &   46.3         &  47.1          \\
SLD (b)              &    26       &   36.3         &  37.9          \\
ALEPH                &    18       &   15.7         &  15.8          \\
OPAL                 &    10       &  112.1 \,      & 110.6 \,       \\
DELPHI (all)         &    23       & \, 5.6         &  \, 6.4        \\
DELPHI (u,d,s)       &    17       & \, 2.1         &  \, 2.4        \\
DELPHI (b)           &    17       &   16.5         &  17.1          \\
\colrule
total                &   249 \,    &  391.2 \,      &  383.2 \,      \\
(/d.o.f.)            &             &  (1.66)        &  (1.62)        \\
\end{tabular*}
\end{ruledtabular}
\end{table}

The difficulty in determining the gluon function is reflected in large
errors of the gluon parameters as shown in Tables \ref{table:parameters-pion},
\ref{table:parameters-kaon}, and \ref{table:parameters-proton}. 
The light-quark ($u$, $\bar u$, $\bar s$) functions also have
large errors which are as large as the gluon ones. The values of
the second moments $M_i^h$ indicate that a large fraction for
the final hadrons comes from the pions. Kaon and
proton contributions are rather small. The moments also
indicate that the favored functions are generally larger than the disfavored
ones as expected. Adding these moments, we find that the total moments are
generally within the energy sum rule in Eq. (\ref{eqn:sum}). However,
the sums slightly exceed one for the gluon and up quark. Considering 
the errors in the moments, we did not strictly impose the sum-rule
condition. 
We found a general tendency that the moment errors for the pion become
smaller in the NLO in comparison with the LO, whereas the NLO errors
are as large as the LO ones for the kaon and proton. Although many used
data are taken at large $Q^2$ as shown in Fig. \ref{fig:zq2}, the pion data
are sensitive to the NLO corrections in the $Q^2$ evolution and coefficient
functions.

Each $\chi^2$ contribution is listed in Tables \ref{tab:chi2-pion},
\ref{tab:chi2-kaon}, and \ref{tab:chi2-proton} for the pion, kaon,
and proton/anti-proton, respectively. 
The $\chi^2$ values indicate that the pion data of
HRS, TOPAZ, SLD (all; $u$,$d$,$s$; $c$), and ALEPH are explained
by our parametrization, whereas
TASSO, TPC, SLD ($b$), OPAL, and DELPHI (all; $u$,$d$,$s$; $b$) data
are not so well reproduced.
In the kaon analysis, the situation is slightly different.
The data of TPC, HRS, TOPAZ, SLD (all; $c$), ALEPH, OPAL, and
DELPHI (all; $u$,$d$,$s$; $b$) are now well explained by our fits,
whereas the data of
TASSO and SLD ($u$,$d$,$s$; $b$) deviate.
In the proton analysis,
the data of TPC, TOPAZ, SLD(all), ALEPH, and
DELPHI (all; $u$,$d$,$s$; $b$) are well explained; however, 
the data of TASSO, HRS, SLD ($u$,$d$,$s$; $c$; $b$), and OPAL
are not reproduced. Some data sets deviate significantly from our
parametrization, namely from other data sets. For example, 
the $\chi^2$ value is more than hundred for ten data points
in the OPAL proton data.

\begin{figure}[t]
\includegraphics[width=0.38\textwidth]{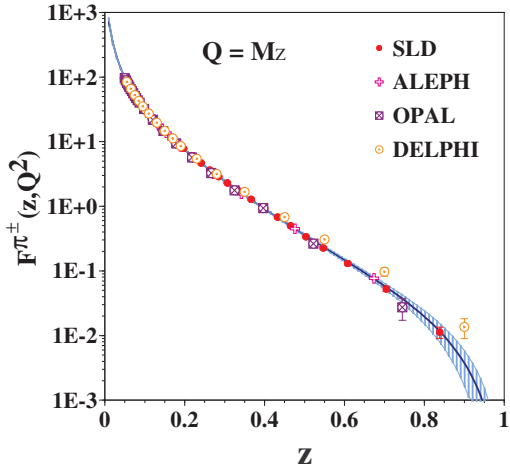}
\vspace{-0.2cm}
\caption{(Color online) Comparison of our NLO results with
        pion-production data at $Q=M_Z$ without separation on initial
        partons by the ALEPH, DELPHI, OPAL, and SLD
        collaborations.}
\label{fig:pion-q-data}
\end{figure}

In Fig. \ref{fig:pion-q-data}, our NLO parametrization results are
compared with the data at $Q=M_Z$ without separation on initial partons
by the ALEPH, DELPHI, OPAL, and SLD collaborations. The figure shows
the comparison with the pion data for $F^{\pi^\pm}(z,Q^2)$ in
Eq. (\ref{eqn:def-ff}). Here, $\pi^\pm$ indicates charged pions:
$\pi^\pm=\pi^+ +\pi^-$. The one-$\sigma$ range of the uncertainty band
is shown by the shaded area. The theoretical curve and its uncertainty
are calculated at $Q=M_Z$. Most of the data are explained by
the parametrization and they are within the uncertainty band.
However, some DELPHI data are outside the band, which leads
to the large $\chi^2$ contribution in Table \ref{tab:chi2-pion}.

\begin{figure}[t!]
\includegraphics[width=0.39\textwidth]{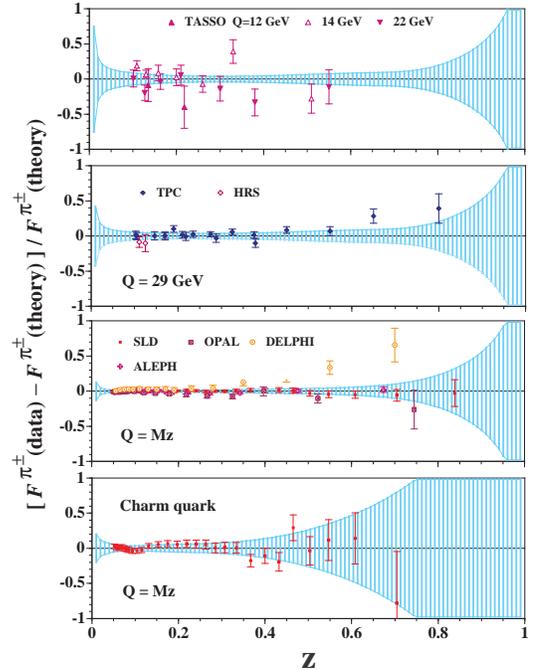}
\vspace{-0.2cm}
\caption{(Color online) Comparison with charged-pion production data
            by the TASSO, TPC, HRS, ALEPH, DELPHI, OPAL, and SLD
            collaborations. The rational differences between
            the data and theoretical calculations are shown
            as a function of $z$. The average scale $Q$=16 GeV is
            taken for theoretical calculations in the top figure with
            the TASSO data.}
\label{fig:pion-data-1}
\end{figure}

\begin{figure}[b!]
\includegraphics[width=0.39\textwidth]{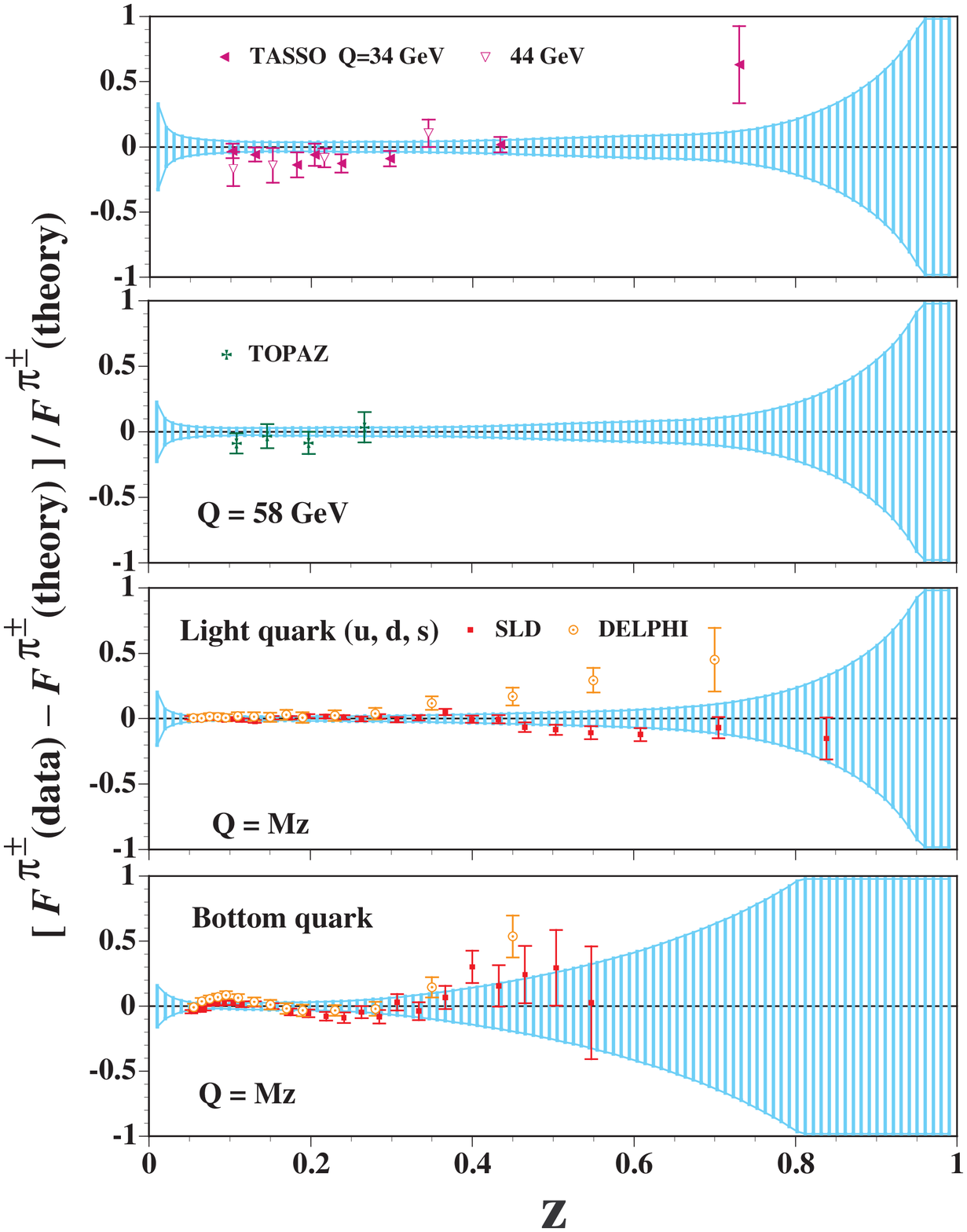}
\vspace{-0.3cm}
\caption{(Color online) Comparison with charged-pion production data
            by the TASSO, TOPAZ, DELPHI, and SLD
            collaborations.  The scale is $Q$=39 GeV
            for theoretical calculations in the top figure.}
\label{fig:pion-data-2}
\end{figure}

Next, the NLO results are compared with each data set for the pion
in Figs. \ref{fig:pion-data-1} and \ref{fig:pion-data-2}, where
the data for the initial-quark separation are also shown.
The figure indicates the rational difference between the data and
the theoretical parametrization:
$[F^{\pi^\pm} ({\rm data})-F^{\pi^\pm} ({\rm theory})]
  /F^{\pi^\pm} ({\rm theory})$
at the same $Q^2$ point with the data.
Most of the data are compatible with the NLO parametrization; however,
some DELPHI data are outside the range of our theoretical
estimations. There may be inconsistency among the data sets. However,
we did not remove the DELPHI data from the analysis because the obtained
functions and the total $\chi^2$ did not change significantly
even if they are excluded from the data set. 
The heavy-quark data also deviate slightly from the theoretical calculations,
which result in the relatively large $\chi^2$ contributions.

\begin{figure}[t!]
\includegraphics[width=0.39\textwidth]{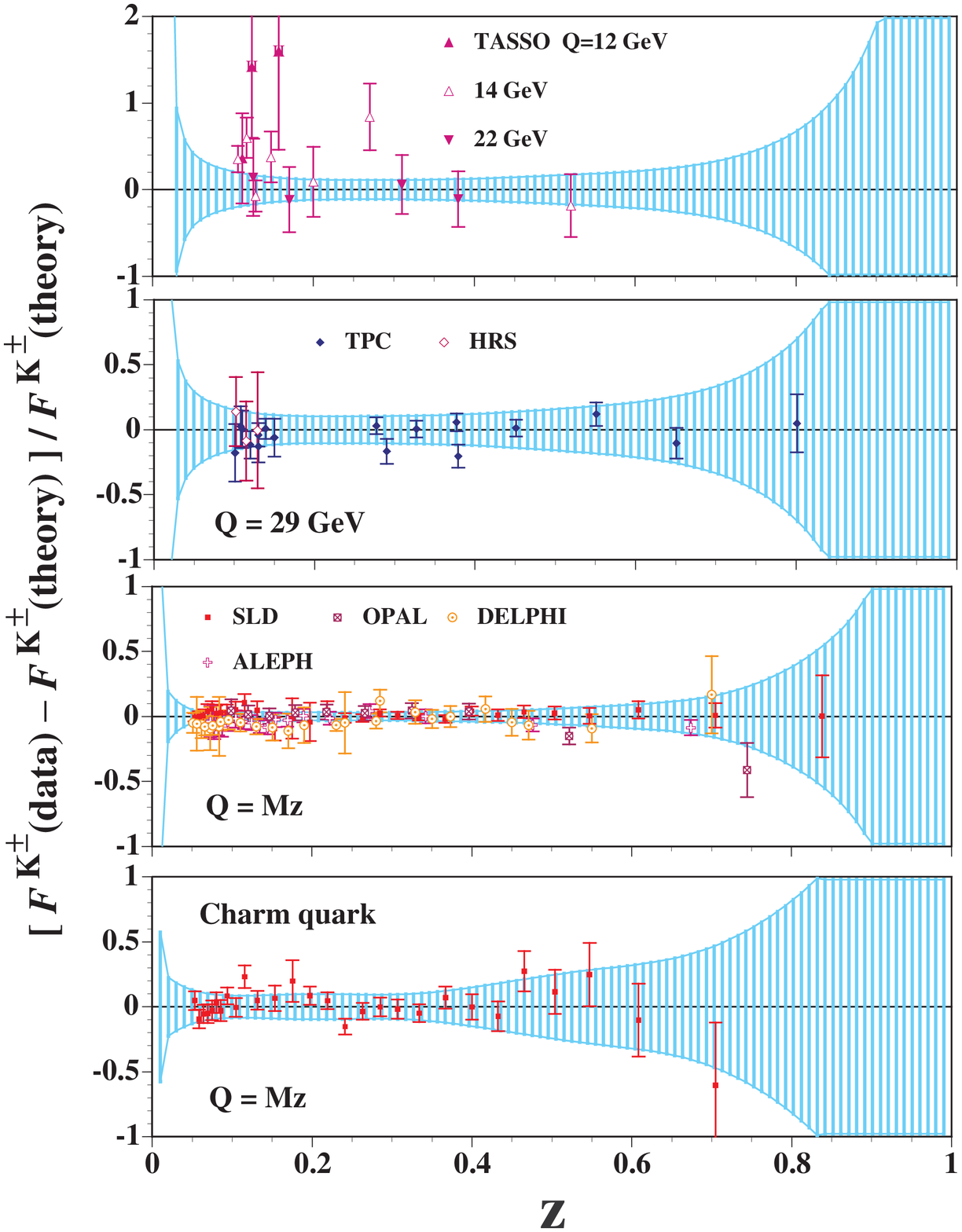}
\vspace{-0.3cm}
\caption{(Color online) Comparison with charged-kaon production data
            by the TASSO, TPC, HRS, ALEPH, DELPHI, OPAL, and SLD
            collaborations. The rational differences between
            the data and theoretical calculations are shown.
            The scale is $Q$=16 GeV for theoretical calculations 
            in the top figure.}
\label{fig:kaon-data-1}
\end{figure}

\begin{figure}[b!]
\includegraphics[width=0.39\textwidth]{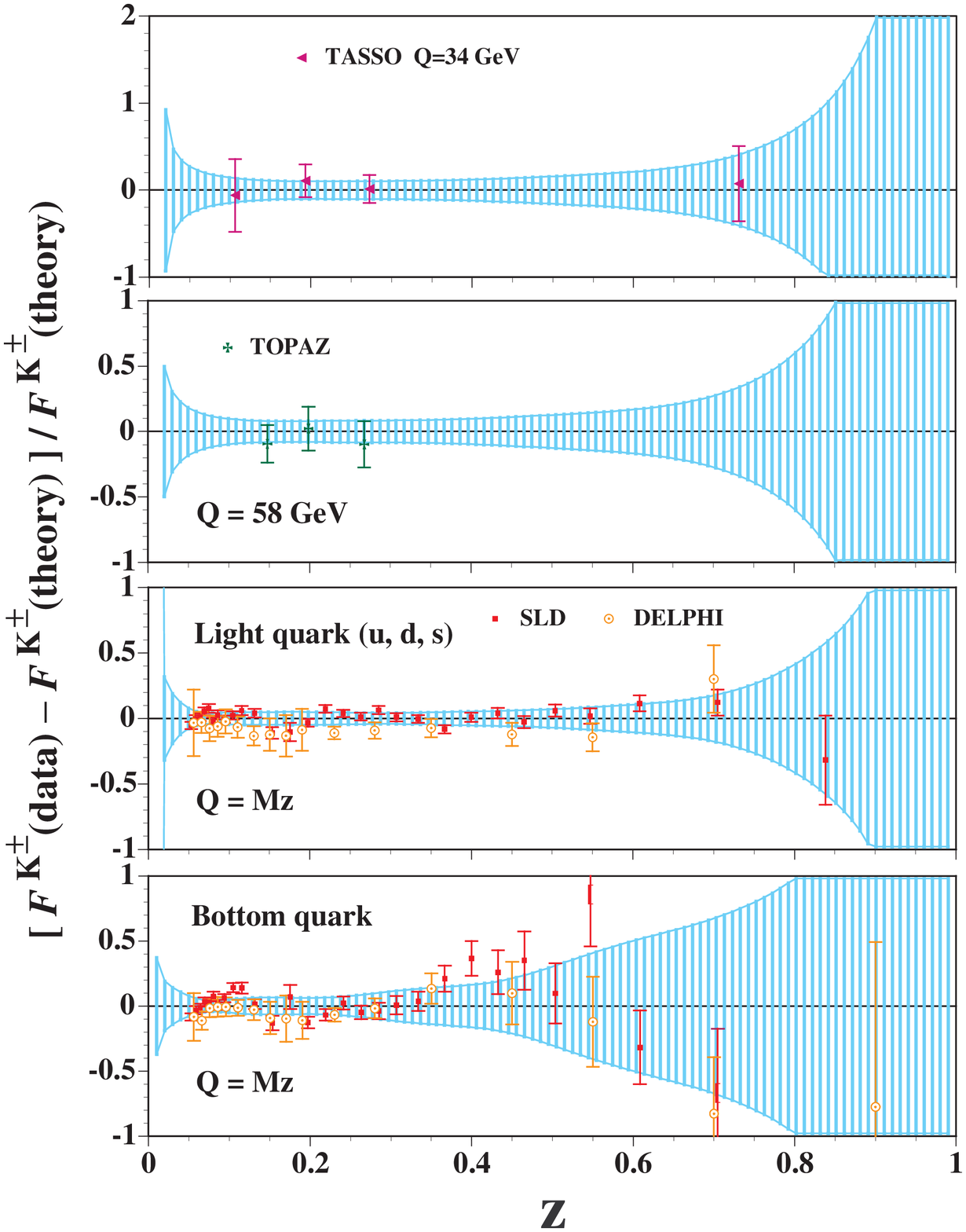}
\vspace{-0.3cm}
\caption{(Color online) Comparison with charged-kaon production data
            by the TASSO, TOPAZ, DELPHI, and SLD collaborations.
            The scale is $Q$=34 GeV for theoretical calculations 
            in the top figure. }
\label{fig:kaon-data-2}
\end{figure}

The kaon fit results are shown in Figs. \ref{fig:kaon-data-1} and
\ref{fig:kaon-data-2}. The kaon data are reproduced
well as shown in these figures. There is no data which indicates
significant deviation from the uncertainty regions, which is reflected
in the smaller $\chi^2$ ($\chi^2$/d.o.f.=1.34 in NLO) than
the value in the pion analysis ($\chi^2$/d.o.f.=1.73).
However, the bottom-quark data by the SLD are not well explained
in the region, $0.4<z<0.5$.

\begin{figure}[t!]
\includegraphics[width=0.39\textwidth]{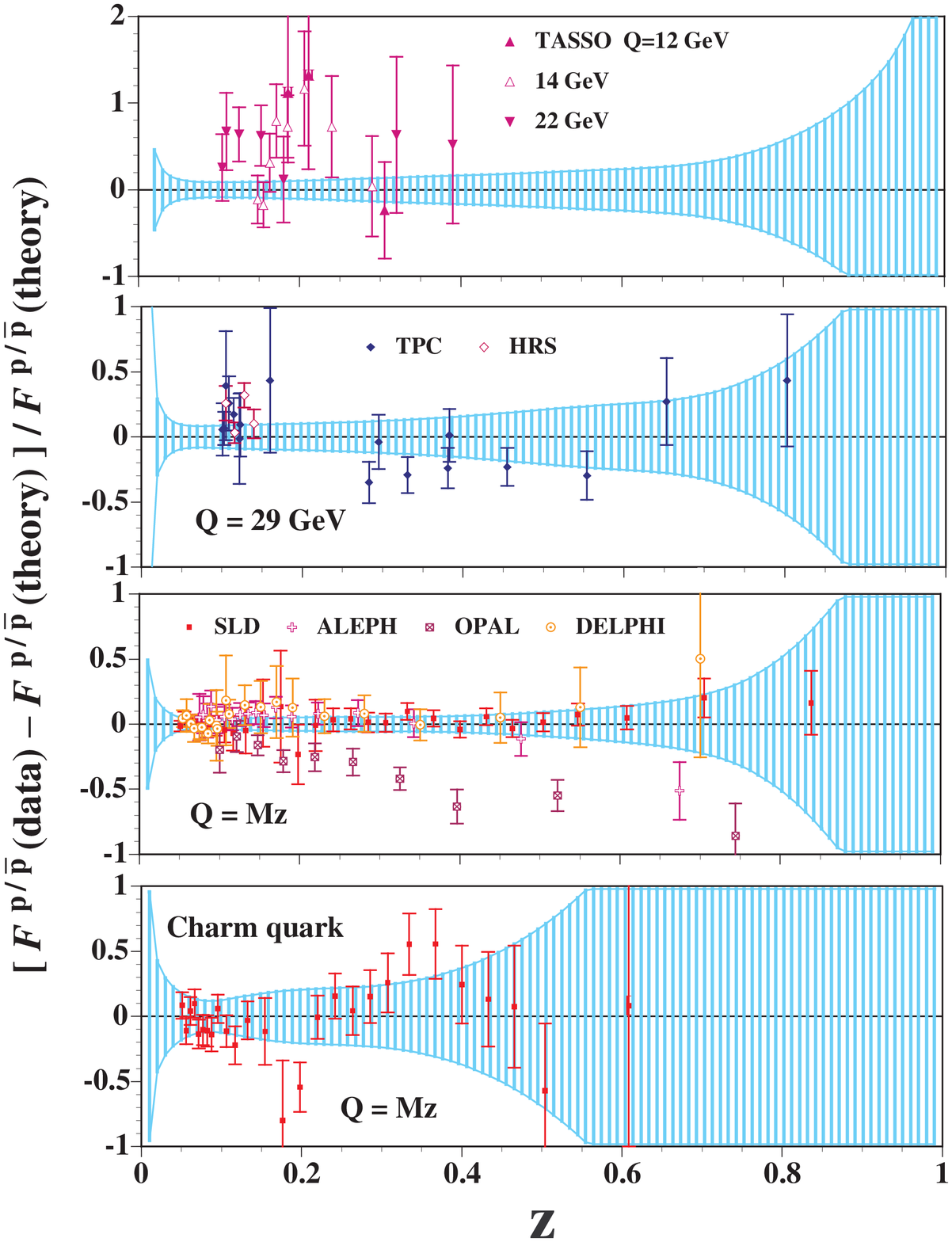}
\vspace{-0.3cm}
\caption{(Color online) Comparison with proton/anti-proton production
            data by the TASSO, TPC, HRS, ALEPH, DELPHI, OPAL, and SLD
            collaborations. The rational differences between
            the data and theoretical calculations are shown.
            The scale is $Q$=16 GeV for theoretical calculations 
            in the top figure.}
\label{fig:proton-data-1}
\end{figure}

\begin{figure}[b!]
\includegraphics[width=0.39\textwidth]{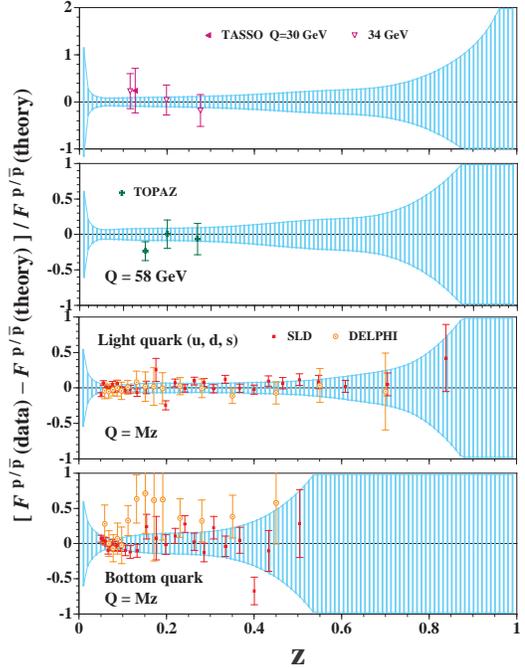}
\vspace{-0.3cm}
\caption{(Color online) Comparison with proton/anti-proton production
            data by the TASSO, TOPAZ, DELPHI, and SLD collaborations.
            The scale is $Q$=32 GeV for theoretical calculations 
            in the top figure.}
\label{fig:proton-data-2}
\end{figure}

The proton results are shown in Figs. \ref{fig:proton-data-1} and
\ref{fig:proton-data-2}. Most of the data are within the uncertainty bands.
However, there are serious deviations for the OPAL data, which results
in the huge $\chi^2$ value ($\chi^2$=110.6 in NLO) for only ten
data points. Since only the OPAL data are different from the other ones
by the ALEPH, DELPHI, and SLD collaborations, there is an inconsistency 
problem for the OPAL measurements.

These figures indicate that our fits are successful.
The uncertainty bands become large at small- and large-$z$ regions,
so that experimental measurements and theoretical studies are needed  
to determine the fragmentation functions in the wide-$z$ range.

\vfill\eject
\subsection{Optimum fragmentation functions and their uncertainties}
\label{ffs}

\begin{figure}[t!]
\includegraphics[width=0.40\textwidth]{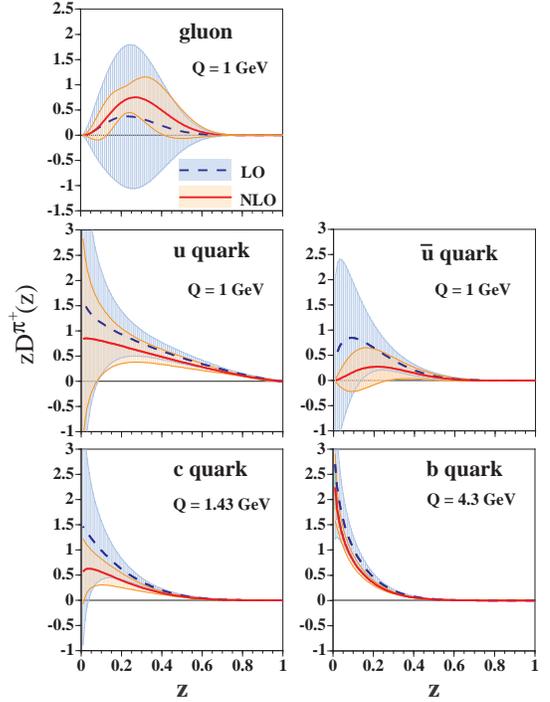}
\vspace{-0.3cm}
\caption{(Color online) Fragmentation functions and their uncertainties
          are shown for $\pi^+$ at $Q^2$=1 GeV$^2$, $m_c^2$, and $m_b^2$.
          The dashed and solid curves indicate LO and NLO results, and
          the LO and NLO uncertainties are shown by the dark- and
          light-shaded bands, respectively.}
\label{fig:pion-ff-q-1}
\end{figure}

\begin{figure}[b!]
\vspace{-0.3cm}
\includegraphics[width=0.40\textwidth]{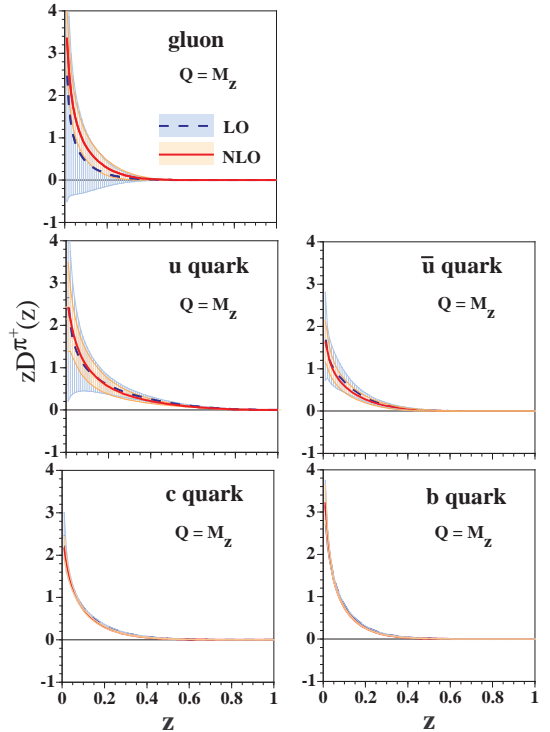}
\vspace{-0.3cm}
\caption{(Color online) Fragmentation functions and their uncertainties
          are shown for $\pi^+$ at $Q^2=M_Z^2$. The notations
          are the same as Fig. \ref{fig:pion-ff-q-1}.}
\label{fig:pion-ff-q-2}
\end{figure}

The obtained fragmentation functions and their uncertainties are shown
for $\pi^+$ in Figs. \ref{fig:pion-ff-q-1} and \ref{fig:pion-ff-q-2}.
The gluon, up-quark, and anti-up-quark functions are shown
at $Q^2$=1 GeV$^2$ in Fig. \ref{fig:pion-ff-q-1}.
The charm- and bottom-quark functions are shown at their thresholds.
The dashed and solid curves indicate LO and NLO results, and
the dark- and light-shaded areas indicate their one-$\sigma$ uncertainty
regions estimated by the Hessian method for the LO and NLO, respectively.
There are differences between the LO and NLO functions.
The gluon function becomes larger in NLO than the LO one,
whereas the quark functions are smaller in NLO.
In the NLO, the favored function $D_u^{\pi^+}$ is the largest, and
the disfavored one $D_{\bar u}^{\pi^+}$ is smaller than $D_u^{\pi^+}$.
The gluon function is in-between, and its moment ($M_g$) is roughly
given by their average, $(M_u+M_{\bar u})/2$, which
agrees with the assumption in Ref. \cite{kretzer}.

Since the experimental data are shown by the sum of light-quark flavors,
the flavor separation as defined by the favored and disfavored initial
functions introduces uncertainties. 
For example, Fig. \ref{fig:pion-data-2} indicates 2\% error
coming directly from the experimental data on the light-quark
($u$, $d$, $s$) fragmentation function at $z=0.2$ and $Q=M_Z$,
whereas the flavor-separated $u$-quark function in the NLO
has 30\% error at $z=0.2$ in Fig. \ref{fig:pion-ff-q-2}.
In order to find such a flavor-separation effect on the uncertainties
at $Q=1$ GeV, the data should be fitted by the function 
$D_{q_s}^h= N_{q_s}^h z^{\alpha_{q_s}^h}(1-z)^{\beta_{q_s}^h}$ where
$q_s=u+\bar u+d+\bar d+s+\bar s$.

An error from the assumed functional
form is not included in estimating the uncertainty bands. 
We fixed one of the gluon parameters ($\beta_g=8$), so
that the uncertainty could be underestimated in $zD_g^{\pi^+}$
at large $z$. However, it would not affect the figure of $zD_g^{\pi^+}$
as long as $\beta_g \gg 1$ because the distribution itself is
small at large $z$.

The errors are large in both LO and NLO, which means that
the fragmentation functions are not well determined particularly
at small $Q^2$. However, it is important to find that all
the functions for the pion are determined much better
in the NLO analysis than the LO ones because the uncertainty
bands are smaller in Fig. \ref{fig:pion-ff-q-1}.
It is especially noteworthy that the gluon function is determined well
in the NLO. The gluonic contributions affect the cross section
through the NLO coefficient function and NLO splitting functions.
Therefore, the shrinkage of the error band suggests that such gluonic
effects are reflected in the current inclusive data for the pion.
In particular, the TASSO collaboration provided many data
in the small $Q^2$ region ($Q^2 << M_Z^2$), and they are important
for identifying such NLO effects in comparison with other data
at $Q^2=M_Z^2$. The reason why the uncertainties are large at small $z$, 
especially in the up and charm functions, is since
small-$z$ data are not included in the analysis.

\begin{figure}[t]
\includegraphics[width=0.40\textwidth]{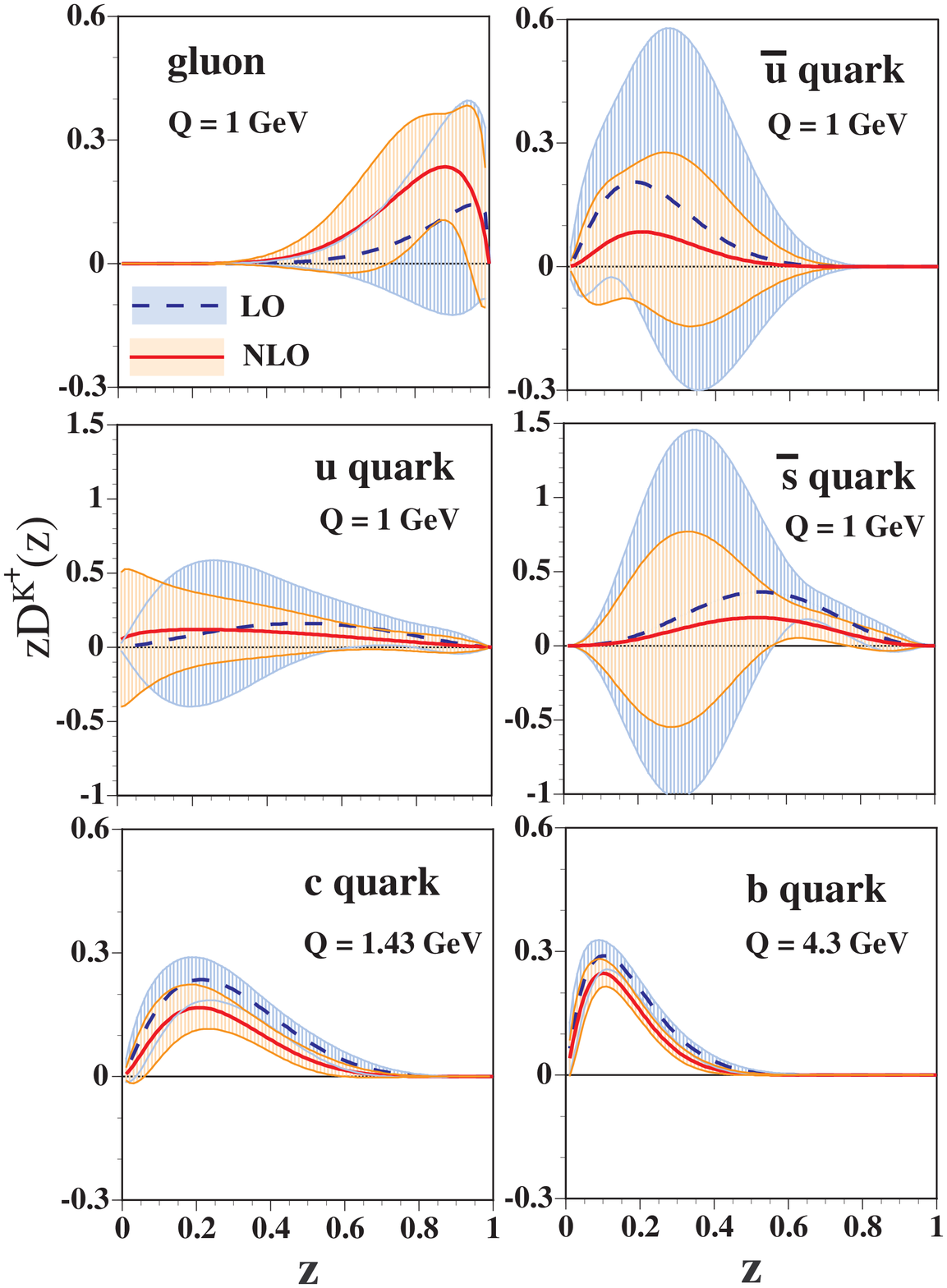}
\vspace{-0.3cm}
\caption{(Color online) Fragmentation functions and their uncertainties
          are shown for $K^+$ at $Q^2$=1 GeV$^2$, $m_c^2$, and $m_b^2$.
          The notations are the same as Fig. \ref{fig:pion-ff-q-1}.}
\label{fig:kaon-ff-q-1}
\end{figure}

\begin{figure}[b!]
\vspace{-0.0cm}
\includegraphics[width=0.40\textwidth]{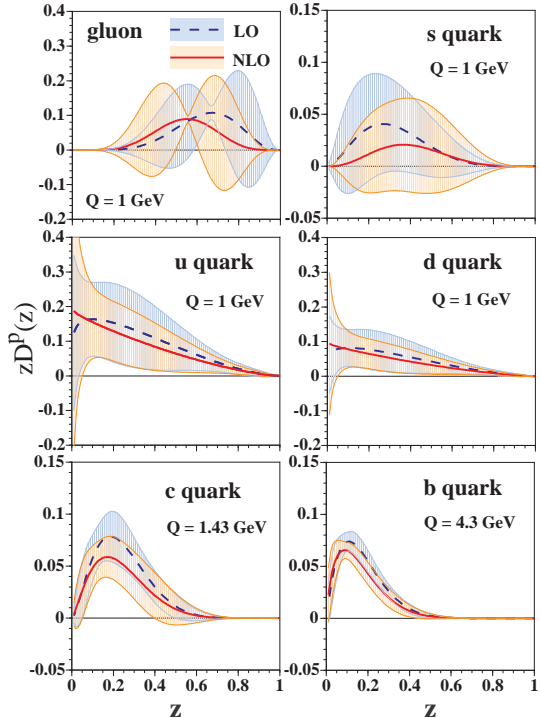}
\vspace{-0.3cm}
\caption{(Color online) Fragmentation functions and their uncertainties
          are shown for the proton at $Q^2$=1 GeV$^2$, $m_c^2$, and $m_b^2$.
          The notations are the same as Fig. \ref{fig:pion-ff-q-1}.}
\label{fig:proton-ff-q-1}
\end{figure}

The functions for the pion are evolved to $Q^2=M_Z^2$, and the results
are shown in Fig. \ref{fig:pion-ff-q-2}. The functions become steep
ones peaked at $z=0$, and the uncertainties become relatively small
in comparison with the ones in Fig. \ref{fig:pion-ff-q-1}.
However, the gluon uncertainties are still large particularly in the LO.
The results in Figs. \ref{fig:pion-ff-q-1} and \ref{fig:pion-ff-q-2}
indicate that the fragmentation functions are not well determined
in the small $Q^2$ region although their uncertainties are
relatively small at large $Q^2$ ($\sim M_Z^2$) especially in the NLO.
Therefore, if the fragmentation functions are used in the small $Q^2$
or small $p_T$ region ($Q^2, p_T^2 << M_Z^2$) such as the lepton-scattering
processes \cite{sidis} of COMPASS, HERMES, and JLab
(Thomas Jefferson National Accelerator Facility) and the hadron collisions
of RHIC \cite{rhic-pi}, it is very important to indicate the uncertainty
ranges of the fragmentation functions.

The obtained fragmentation functions and their uncertainties are shown
for $K^+$ in Fig. \ref{fig:kaon-ff-q-1} at $Q^2$=1 GeV$^2$,
$m_c^2$, and $m_b^2$. There are similar tendencies to the pion results.
The obtained gluon function in the NLO is larger than the LO one
as the pion case, whereas the quark functions are smaller.
It is interesting to find that the anti-strange function
is generally larger than the up function: $D_{\bar s}^{K^+} > D_{u}^{K^+}$,
which could be interpreted in the following way. In order to create
$K^+$ from a parent $\bar s$ (or $u$), a $u\bar u$ ($s\bar s$) pair
needs to be created. Since the strange-quark mass is larger, the $s\bar s$
creation could be suppressed in comparison with the $u\bar u$ creation,
which leads to the inequality. However, the large uncertainty bands
indicate that the separation between $u$ and $\bar s$ functions 
is difficult.

There is a conspicuous difference between the gluon functions
for the pion and kaon. The gluon function ($zD_g^{K^+}$) is peaked
at large $z$, whereas it is at $z=0.2 \sim 0.3$ in the pion.
Even if an initial distribution with a peak at small $z$ is 
supplied in the $\chi^2$ fit, the outcome is always peaked
at large $z$. It could be physically understood in the following simple
picture. In order to produce $K^+$ from a gluon, the gluon should first
split into a $s\bar s$ pair. Then, another gluon is emitted from the $s$
or $\bar s$ quark, and it subsequently splits into a $u\bar u$ pair.
It requires higher energy for the parent gluon to produce the $s\bar s$
pair ($g \rightarrow s\bar s$) in the kaon creation than the one
for a $u \bar u$ pair ($g \rightarrow u\bar u$) 
or $d \bar d$ pair ($g \rightarrow d\bar d$) in the pion creation
because of the mass difference. The higher energy means that the function
is peaked at larger $z$ in the kaon. 

The kaon functions also have large uncertainties in both
favored and disfavored cases. They have slightly larger errors than
the pionic ones if the ratios $\delta D_i /D_i$ are considered.
The uncertainty bands become smaller in NLO than the LO ones.
However, the NLO improvement is not as clear as the pionic one.
A possible reason is that many accurate data are not taken at small
$Q^2$ ($<<M_Z^2$), for example, by the TASSO collaboration as for
the pion.

The fragmentation functions for the proton are shown in
Fig. \ref{fig:proton-ff-q-1} at $Q^2$=1 GeV$^2$, $m_c^2$, and $m_b^2$.
Here, the gluon moments are fixed by the favored and disfavored moments,
so that they are almost the same in LO and NLO. 
As expected, the favored functions $D_u^p$ and $D_d^p$ are larger
than the disfavored functions. The gluon functions have peaks
in the medium-$z$ region. In general, the proton functions are also
not determined well, and the uncertainties are as large as the kaonic
ones. The NLO improvement is also not obvious in the proton.
This fact suggests that the current proton and antiproton data
should not be much sensitive to the NLO corrections.

Since the gluon moment is given by the average of favored
and disfavored moments, the error of the gluon function $D_g^p$
could be underestimated. A noticeable difference from
the pion and kaon figures is that the gluon uncertainty
bands shrink in the region $0.5<z<0.7$, which is caused
by fixing the moment. There are large
contributions from the diagonal terms in Eq. (\ref{eqn:ddih}). 
The term $(\partial D_g / \partial M_g)^2$ is a smooth function of $z$
with a peak in the same position as the function $D_g$.
On the other hand, $(\partial D_g / \partial \alpha_g)^2$ and
$(\partial D_g / \partial \beta_g)^2$ have double peaks,
which come from destructive interferences between two derivative
terms. For example, the parameter $N_g$ depends on the parameter 
$\beta_g$ by Eq. (\ref{eqn:nih}). The term $\partial N_g/\partial \beta_g$
is positive, whereas there is a negative contribution from the
derivative of $(1-z)^{\beta_g}$.
In the pion case,
$(\partial D_g / \partial M_g)^2 H_{M_g M_g}^{-1}$
is several times larger than 
$(\partial D_g / \partial \alpha_g)^2 H_{\alpha_g \alpha_g}^{-1}$,
so that such double-peak structure does not appear 
in the uncertainty band in Fig. \ref{fig:pion-ff-q-1}.
However, the term
$(\partial D_g / \partial M_g)^2 H_{M_g M_g}^{-1}$
does not exist in the proton analysis since the parameter $M_g$
is fixed. Therefore, the double-peak shape of 
$(\partial D_g / \partial \beta_g)^2 H_{\beta_g \beta_g}^{-1}$
becomes apparent in Fig. \ref{fig:proton-ff-q-1}.

The determined fragmentation functions and their uncertainties can be
calculated by using a library code, which is obtained from 
our web page \cite{ff-web-kek}. The details are explained 
on the web page and distributed files.

In these analyses for the pion, kaon, and proton, the charm and bottom
functions are determined mainly from the heavy-quark tagging data
by the SLD and DELPHI collaborations. The light-quark functions are
constrained especially by the accurate data of the SLD. 
In order to determine the fragmentation functions accurately,
we need low-energy data possibly by the Belle and Babar collaborations.
The low-energy data should be important for the determination
of the gluon functions $D_g^h$ which are essential for describing
the low-$p_T$ hadron productions, for example, at RHIC. 

The fragmentation functions for $\pi^-$, $K^-$, and $\bar p$ can be
calculated from the obtained functions by using the relations in
Eqs.  (\ref{eqn:pi-k-pbar-q}) and (\ref{eqn:pi-k-pbar-g}).
The fragmentation functions for $\pi^0$, $K^0$, $\bar K^0$, $n$, and $\bar n$
can be calculated by using the determined functions for the $\pi^+$, $K^+$,
and proton. The relations are explained in Appendix \ref{other-hadrons}.

\subsection{Comparison with other parametrizations}
\label{other-ffs}

\begin{figure}[t]
\includegraphics[width=0.37\textwidth]{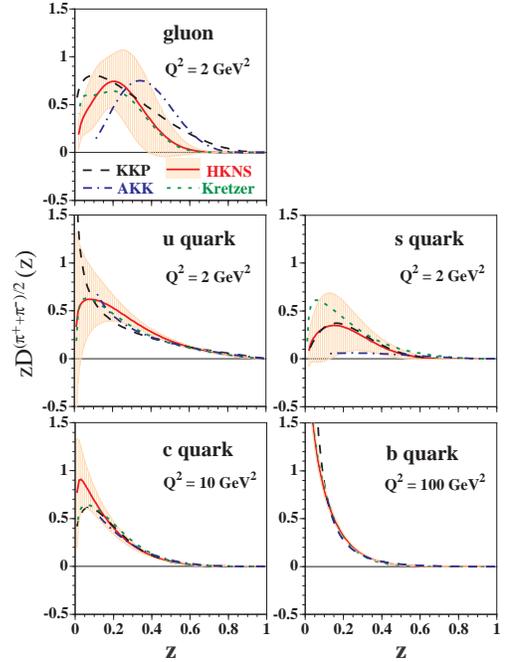}
\vspace{-0.3cm}
\caption{(Color online) The NLO fragmentation functions 
          for the pion, $z(D_i^{\pi^+}+D_i^{\pi^-})/2$,
          are compared with other parametrizations 
          at $Q^2$=2 GeV$^2$, 10 GeV$^2$, and 100 GeV$^2$.
          The solid, dashed, dash-dotted, and dotted curves indicate
          HKNS, KKP, AKK, and Kretzer parametrizations, respectively,
          in the NLO $\overline {\rm MS}$ scheme. The HKNS uncertainty
          bands are shown by the shaded areas.}
\label{fig:pion-ff-comp}
\end{figure}

\begin{figure}[b!]
\includegraphics[width=0.37\textwidth]{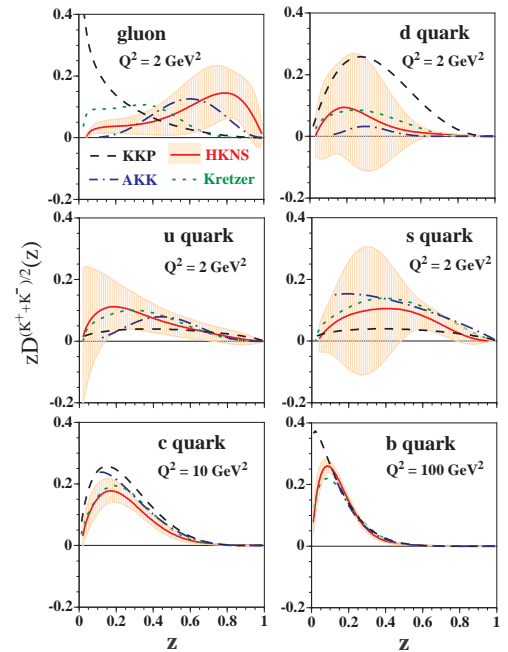}
\vspace{-0.3cm}
\caption{(Color online) The NLO fragmentation functions 
          for the kaon, $z(D_i^{K^+}+D_i^{K^-})/2$,
          are compared with other parametrizations 
          at $Q^2$=2 GeV$^2$, 10 GeV$^2$, and 100 GeV$^2$.
          The notations are the same as Fig. \ref{fig:pion-ff-comp}.}
\label{fig:kaon-ff-comp}
\end{figure}

\begin{figure}[t]
\vspace{-0.1cm}
\includegraphics[width=0.37\textwidth]{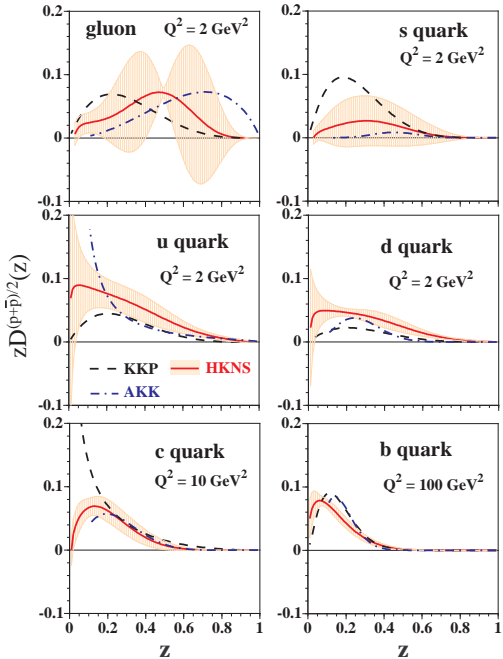}
\vspace{-0.3cm}
\caption{(Color online) The NLO fragmentation functions 
          for the proton/anti-proton, $z(D_i^{p}+D_i^{\bar p})/2$,
          are compared with other parametrizations 
          at $Q^2$=2 GeV$^2$, 10 GeV$^2$, and 100 GeV$^2$.
          The notations are the same as Fig. \ref{fig:pion-ff-comp}.}
\label{fig:proton-ff-comp}
\end{figure}

Our NLO fragmentation functions are compared with other parametrizations
by KKP \cite{kkp}, AKK \cite{kkp}, and Kretzer \cite{kretzer}
in Figs. \ref{fig:pion-ff-comp}, \ref{fig:kaon-ff-comp},
and \ref{fig:proton-ff-comp} for the pion, kaon, and proton/anti-proton,
respectively. Since $\pi^+$, $K^+$, and proton are not separated
from $\pi^-$, $K^-$, and anti-proton in the KKP and AKK parametrizations,
the combinations $z(D_i^{\pi^+}+D_i^{\pi^-})/2$, 
$z(D_i^{K^+}+D_i^{K^-})/2$, and $z(D_i^{p}+D_i^{\bar p})/2$ are shown.
Our parametrization is denoted HKNS (Hirai, Kumano, Nagai, Sudoh) in
these figures. 
The light-quark functions are separated in the AKK parametrization
due to additional OPAL data, which are not used in the other analyses.
The KKP, AKK, and Kretzer distributions are calculated
by using their library codes. The small-$z$ part is not plotted
for the AKK because their code does not support the region, $z<0.1045$
by considering resummation effects in comparison with the data.
The heavy-quark thresholds are taken $4m_c^2$ and $4m_b^2$
for the fragmentation functions in the KKP and AKK parametrizations,
whereas they are $m_c^2$ and $m_b^2$ in the Kretzer and HKNS analyses.
The minimum value of $Q^2$ in the KKP and AKK codes is $Q^2$=2 GeV$^2$.
Therefore, comparisons are made at $Q^2$=2, 10, and 100 GeV$^2$. 

If the distributions are compared with each other in Fig. \ref{fig:pion-ff-comp}
for the pion, they agree well in up-, charm-, and bottom-quark functions. 
The up-quark function of the KKP has a singular behavior as $z \rightarrow 0$,
which is different from the Kretzer's, AKK, and HKNS functions. However,
the small-$z$ data are not included in these analyses, so that
the differences should not be taken seriously at small $z$.
Since new SLD data in 2004 are accurate and they are not
used in other parametrizations, the fragmentation functions
could be better determined in our parametrization.
The four types of functions are very different in the gluon
and disfavored strange-quark functions in Fig. \ref{fig:pion-ff-comp}.
All the gluon functions have various peak positions in the region
$0.1<z<0.4$ and functional forms are different.
Our gluon function roughly agrees with the Kretzer's function.
The four strange-quark functions are also much different. Although
our strange-quark function is almost equal to the KKP, the Kretzer's (AKK)
function is much larger (smaller). However, it is important to find that
the gluon and strange-quark functions, needless to say the up-, charm-, and
bottom-quark functions, by the KKP, AKK, Kretzer, and HKNS are
consistent with each other because they are within the uncertainty
bands. This fact indicates that all the analyses are successful and 
consistent. 

For the kaon, the KKP, AKK, Kretzer's, and HKNS results
agree well in the charm- and bottom-quark functions
as shown in Fig. \ref{fig:kaon-ff-comp}. However,
they differ much in the gluon and light-quark functions.
In particular, the HKNS and AKK gluon functions are peaked at large $z$,
whereas the KKP diverges at small $z$ and the Kretzer's function is peaked
at $z \sim 0.4$. The KKP and Kretzer functions are outside
the uncertainty band at small $z$. Since the parameter $\alpha_g$ is fixed
and because of the functional form with the peak at large $z$,
the uncertainty could be underestimated in the small-$z$ region. 
The large variations among the parametrizations
of $D_g^{(K^++K^-)/2}$ indicate that the current kaon data are not
accurate enough to fix the gluon fragmentation function for the kaon from 
the scaling violation. Lower-energy date, for example from Belle and BaBar,
should be able to improve the situation. The light-quark ($u$, $d$, $s$)
functions are also very different; however, they are roughly within
the uncertainty band. Except for the gluon function, all
the analysis results are consistent in the kaon.

The results agree well with each other in the charm- and
bottom-quark functions for the proton and anti-proton as shown in
Fig. \ref{fig:proton-ff-comp}, although there are some differences
at small $z$ where experimental data are not used in the analyses.
The Kretzer's parametrization is not available for the proton.
The three gluon functions are very different but they are roughly
within the uncertainty band. It is almost impossible to determine
the accurate gluon function at this stage. The disfavored strange-quark
functions are also very different. In particular, the KKP (AKK) function
is much larger (smaller) than our result. Our favored functions,
up- and down-quark ones, are larger than the KKP and AKK functions.
However, all these fragmentation functions including the disfavored
ones are roughly within the estimated uncertainties.

From these comparisons, we found that the analyses of KKP, AKK, Kretzer,
and HKNS are consistent because they are generally within the uncertainty
bands estimated by the Hessian method. However, there are noticeable 
differences in the gluon and light-quark fragmentation functions. 
They should be clarified by future measurements especially at small $Q^2$.
The large uncertainties could cause serious effects in discussing
hadron-production processes at small $p_T$ such as RHIC, HERMES, and 
JLab.

\section{\label{summary} Summary}

Unpolarized fragmentation functions for the pion, kaon, and proton
have been determined in the LO and NLO from global analyses of
$e^+ +e^- \rightarrow h+X$ data. Their uncertainties
were estimated by the Hessian method. We found the large uncertainties
in the fragmentation functions at small $Q^2$ ($\sim$1 GeV$^2$) although
they become relatively smaller at high energies ($Q^2 \sim M_Z^2$).
In particular, the gluon and light-quark fragmentation functions have
large uncertainties.
However, they are determined more accurately in the NLO analyses
than the LO ones for the pion and kaon. Because of the large uncertainties
at small $Q^2$, it is important that such errors need to be taken into
account in analyzing the hadron-production data in high-energy
lepton-proton, proton-proton, and nuclear reactions. 
Low-energy $e^+ e^-$ measurements should be valuable for determining
especially the gluon fragmentation functions by the scaling violation.
A code for calculating the determined fragmentation functions can be
obtained from our web page \cite{ff-web-kek}.

\begin{acknowledgements}
S.K. and M.H. were supported by the Grant-in-Aid for Scientific Research from
the Japanese Ministry of Education, Culture, Sports, Science, and Technology.
T.-H.N. was supported by the JSPS Research Fellowships for Young Scientists.
S.K., T.-H.N., and K.S. thank Institute for Nuclear Theory at the University
of Washington for its hospitality and the US Department of Energy for partial
support.
\end{acknowledgements}

\appendix
\section{Fragmentation functions for
$\pi^0$, $K^0$, $\bar K^0$, $n$, and $\bar n$}
\label{other-hadrons}

The fragmentation functions of $\pi^+$, $\pi^-$, $K^+$, $K^-$, $p$, and
$\bar p$ are extracted from the experiment by assuming charge symmetry
in relating $D_i^{\pi^+}$, $D_i^{K^+}$, and $D_i^{p}$ to 
the corresponding ones $D_i^{\pi^-}$, $D_i^{K^-}$, and $D_i^{\bar p}$
by Eqs. (\ref{eqn:pi-k-pbar-q}) and (\ref{eqn:pi-k-pbar-g}).
One may need the fragmentation functions for $\pi^0$, $K^0$, $\bar K^0$,
$n$, or $\bar n$ in one's studies of high-energy hadron reactions.
The quark compositions of these hadrons are 
$\pi^0 ((u\bar u - d\bar d)/\sqrt{2})$, $K^0(d\bar s)$, 
$\bar K^0(\bar d s)$, $n \, (udd)$, and $\bar n \, (\bar u \bar d \bar d)$.
Considering these compositions, we relate the obtained functions to
the ones for $\pi^0$, $K^0$, $\bar K^0$, $n$, and $\bar n$ \cite{kkp}.
The $\pi^0$ functions are given by the averages of $\pi^+$ and $\pi^-$:
\begin{equation}
D_{i}^{\pi^0} (z,Q^2) = \frac{1}{2} [ \, D_{i}^{\pi^+} (z,Q^2)
                                        +D_{i}^{\pi^-} (z,Q^2) \, ] .
\label{eqn:pi0-ff}
\end{equation}
The $K^0$ functions are also calculated by the relations:
\begin{alignat}{2}
D_{d}^{K^0} (z,Q^2) & = D_{u}^{K^+} (z,Q^2) , & &
\nonumber \\
D_{\bar s}^{K^0} (z,Q^2) & = D_{\bar s}^{K^+} (z,Q^2) , & &
\nonumber \\
D_{u}^{K^0} (z,Q_0^2) & = D_{\bar u}^{K^+} (z,Q_0^2) 
                      & = \, & D_{\bar u}^{K^0} (z,Q_0^2) 
\nonumber \\
                      & = D_{\bar d}^{K^0} (z,Q_0^2) 
                      & = \, & D_{s}^{K^0} (z,Q_0^2) ,
\nonumber \\        
D_{c,b}^{K^0} (z,Q^2) & = D_{c,b}^{K^+} (z,Q^2) &
                 = \, & D_{\bar c,\bar b}^{K^0} (z,Q^2) , 
\nonumber \\
D_{g}^{K^0} (z,Q^2) & = D_{g}^{K^+} (z,Q^2) . & & 
\label{eqn:k0-ff}
\end{alignat}
However, one should be careful about the following relations
at different $Q^2$ from $Q_0^2$ because of the NLO evolution:
\begin{equation}
D_{\bar u}^{K^0} (z,Q^2) \ne \, D_{\bar d}^{K^0} (z,Q^2)
                         \ne \, D_{s}^{K^0} (z,Q^2) ,
\label{eqn:k0-ff-2}
\end{equation}
although $D_{u}^{K^0} (z,Q^2) = D_{\bar u}^{K^0} (z,Q^2)$ is still
valid. Then, $\bar K^0$ functions are related to the $K^0$ ones by
\begin{align}
D_{q}^{\bar K^0} (z,Q^2) & = D_{\bar q}^{K^0} (z,Q^2) ,
\nonumber \\
D_{g}^{\bar K^0} (z,Q^2) & = D_{g}^{K^0} (z,Q^2) .
\label{eqn:k0bar-ff}
\end{align}
The neutron functions are related to the proton ones by
\begin{align}
D_{u}^{n} (z,Q^2) & = D_{d}^{p} (z,Q^2) ,
\nonumber \\
D_{d}^{n} (z,Q^2) & = D_{u}^{p} (z,Q^2) ,
\nonumber \\
D_{i}^{n} (z,Q^2) & = D_{i}^{p} (z,Q^2) 
                     \ \ \text{for $i \ne u, \, d$} .
\label{eqn:n-ff}
\end{align}
The anti-neutron functions are then given by
\begin{align}
D_{q}^{\bar n} (z,Q^2) & = D_{\bar q}^{n} (z,Q^2) ,
\nonumber \\
D_{g}^{\bar n} (z,Q^2) & = D_{g}^{n} (z,Q^2) .
\label{eqn:nbar-ff}
\end{align}
Using the code in Ref. \cite{ff-web-kek}, one should be able calculate
the fragmentation functions for
$\pi^+$, $\pi^0$, $\pi^-$, $K^+$, $K^-$, $K^0$, $\bar K^0$, $p$, $\bar p$,
$n$, and $\bar n$ at given $z$ and $Q^2$.



\end{document}